\documentclass{aa}  

\usepackage{natbib}

\usepackage{caption}
\usepackage{subcaption}

\usepackage[flushleft]{threeparttable}

\usepackage{graphicx}
\usepackage{txfonts}
\usepackage{gensymb}

\usepackage{url}

\makeatletter
\newcommand\footnoteref[1]{\protected@xdef\@thefnmark{\ref{#1}}\@footnotemark}
\makeatother

\begin{document} 

   \title{LOFAR Deep Fields: Probing a broader population of polarized radio galaxies in ELAIS-N1}

%   \subtitle{I. Overviewing the $\kappa$-mechanism}

   \author{N. Herrera Ruiz
          \inst{1}
          \and
          S. P. O'Sullivan\inst{2}
          \and
          V. Vacca\inst{3}
          \and
          V. Jeli\'c \inst{4}
          \and
          B. Nikiel-Wroczy\'nski \inst{5}
          \and
          S. Bourke\inst{6}
          \and
          J. Sabater\inst{7}          
          \and
          R.-J. Dettmar\inst{1}
          \and
          G. Heald\inst{8}
          \and 
          C. Horellou\inst{6}
          \and
          S. Piras\inst{6}
          \and
          C. Sobey\inst{8}    
          \and
          T. W. Shimwell\inst{9}
          \and
          C. Tasse\inst{10,11,12}
          \and
          M. J. Hardcastle\inst{13}   
          \and
          R. Kondapally\inst{7}
          \and
          K. T. Chy\.zy\inst{5}         
          \and 
          M. Iacobelli\inst{9}          
          \and
          P. N. Best\inst{7}
          \and 
          M. Br\"uggen\inst{14}
          \and
          E. Carretti\inst{15}
          \and
          I. Prandoni\inst{15}
          }

   \institute{Ruhr University Bochum, Faculty of Physics and Astronomy, Astronomical Institute, Universit\"atstrasse 150, 44801 Bochum, Germany\\
              \email{herrera@astro.rub.de}
   \and School of Physical Sciences and Centre for Astrophysics \& Relativity, Dublin City University, Glasnevin, D09 W6Y4, Ireland
   \and INAF – Osservatorio Astronomico di Cagliari, Via della Scienza 5, I-09047 Selargius (CA), Italy
   \and Ru\dj{}er Bo{\v s}kovi\'c Institute, Bijeni{\v c}ka cesta 54, 10000  Zagreb, Croatia
   \and Astronomical Observatory, Jagiellonian University, ul. Orla 171, 30-244, Krak\'ow, Poland
   \and Department of Space, Earth and Environment, Chalmers University of Technology, Onsala Space Observatory, 439 92 Onsala, Sweden
   \and SUPA, Institute for Astronomy, Royal Observatory, Blackford Hill, Edinburgh, EH9 3HJ, UK
   \and CSIRO Astronomy and Space Science, PO Box 1130, Bentley, WA, 6102, Australia
   \and ASTRON, the Netherlands Institute for Radio Astronomy, Postbus 2, 7990 AA, Dwingeloo, The Netherlands
   \and GEPI, Observatoire de Paris, CNRS, Université Paris Diderot, 5 place Jules Janssen, 92190 Meudon, France
   \and Centre for Radio Astronomy Techniques and Technologies, Department of Physics and Electronics, Rhodes University, Grahamstown 6140, South Africa
   \and USN, Observatoire de Paris, CNRS, PSL, UO, Nançay, France
   \and Centre for Astrophysics Research, University of Hertfordshire, College Lane, Hatfield AL10 9AB, UK
   \and Hamburger Sternwarte, Universit\"at Hamburg, Gojenbergsweg 112, 21029 Hamburg, Germany
   \and INAF-IRA, Via P. Gobetti 101, 40129 Bologna, Italy 
}

   \date{Received July 13, 2020; accepted August 25, 2020} %September 15, 1996; accepted March 16, 1997}

  \abstract{We present deep polarimetric observations of the European Large Area {\it ISO} Survey-North\,\,1 (ELAIS-N1) field using the Low Frequency Array (LOFAR) at 114.9-177.4\,MHz. The ELAIS-N1 field is part of the LOFAR Two-metre Sky Survey deep fields data release I. For six eight-hour observing epochs, we align the polarization angles and stack the 20$''$-resolution Stokes $Q$, $U$-parameter data cubes. This produces a 16~deg$^{2}$ image with 1$\sigma_{\rm QU}$ sensitivity of 26~$\mu$Jy~beam$^{-1}$ in the central area. In this paper, we demonstrate the feasibility of the stacking technique, and we generate a catalog of polarized sources in ELAIS-N1 and their associated Faraday rotation measures (RMs). While in a single-epoch observation we detect three polarized sources, this number increases by a factor of about three when we consider the stacked data, with a total of ten sources. This yields a surface density of polarized sources of one per 1.6~deg$^{2}$. The Stokes $I$ images of three of the ten detected polarized sources have morphologies resembling those of FR~I radio galaxies. This represents a greater fraction of this type of source than previously found, which suggests that more sensitive observations may help with their detection.}

   \keywords{polarization --
                galaxies: individual (ELAIS-N1) --
                radio continuum: galaxies  --  
                techniques: polarimetric             
               }

   \titlerunning{LOFAR Deep Fields: polarized radio galaxies in ELAIS-N1}
   \authorrunning{N. Herrera Ruiz et al.}

   \maketitle

\section{Introduction}
\label{sec:int}

The Low Frequency Array (LOFAR, \citealt{vanhaarlem2013}) is a phased array radio interferometer, developed to explore the low-frequency radio sky (below 250~MHz). Because of its large field of view (>10~deg$^2$ at 150~MHz), LOFAR is ideal for performing large-area sky surveys. 

In addition to being a general purpose facility, LOFAR has been used to conduct several large key projects. These include studies of the Epoch of Reionization (EoR), deep extragalactic surveys, transient sources, ultra high energy cosmic rays, solar science and space weather, and cosmic magnetism.

The present project is carried out within the framework of the LOFAR Magnetism Key Science Project (MKSP\footnote{http://lofar-mksp.org}), which aims to study the magnetized Universe using LOFAR. Magnetic fields are ubiquitous throughout the Universe, and they play an important role in galaxy and galaxy cluster evolution as well as in the interstellar and intracluster media (e.g., \citealt{beck2013}). The Faraday rotation measure (RM) quantifies the degree of rotation undergone by the polarization position angle, as a function of wavelength-squared, due to magneto-ionic media between the emitting source and the observer. Therefore, the RM provides information about magnetic fields along the line of sight (e.g., \citealt{Han2017}). Catalogs of RMs of polarized radio sources can be used to study the Faraday rotation of our Galaxy and constrain its large-scale magnetic fields (e.g., \citealt{hutschenreuter2020,vaneck2011,sobey2019}). These catalogs are also powerful tools for the study of extragalactic large-scale magnetic fields, as shown by \citet{bonafede2010}, who determined the Coma cluster magnetic field strength from RMs. Moreover, due to the high sensitivity and high precision of  new-generation radio interferometers, RM catalogs of polarized radio sources will be key components for the investigation of the magnetization of the cosmic web (e.g., \citealt{vacca2016, osullivan2020}). 

Polarization observations can be used to investigate the emitting sources and the magneto-ionic media moving toward them by using Faraday rotation and frequency-dependent depolarization measurements (e.g., \citealt{burn1966, sokoloff1998}). For example, different source populations can be distinguished and characterized \citep{farnes2014, anderson2015}. Two of the main science goals that are particularly relevant to deep field polarization observations are: i) the study of galaxy evolution through the analysis of polarized source populations and the characterization of individual polarized sources; and ii) to enlarge the database of extragalactic Faraday RMs used for the study of cosmic magnetism. Here, we present a catalog of polarized radio sources detected using LOFAR observations of the European Large Area ISO Survey-North 1 (ELAIS-N1) field, including their RMs and additional characteristics. 

The ELAIS-N1 field was one of the regions observed by the ELAIS survey \citep{oliver2000} in the Northern Hemisphere (RA\,=\,16$^{h}$10$^{m}$01$^{s}$,  Dec\,=\,54\degree30$^{\prime}$36$^{\prime\prime}$). This survey was a single Open Time project conducted by the Infrared Space Observatory (ISO, \citealt{kessler1996}). This field has exceptional multiwavelength coverage. It has been observed at X-ray \citep{manners2003}, ultraviolet \citep{martin2005,morrissey2007}, optical \citep{mcmahon2001,aihara2018}, and infrared wavelengths \citep{lawrence2007,lonsdale2003,mauduit2012,oliver2012}. \citet{kondapally2020} present a detailed description of the multiwavelength coverage of the field. At radio frequencies, the ELAIS-N1 field was observed with the Giant Metrewave Radio Telescope (GMRT) at 325\,MHz \citep{sirothia2009} and at 610\,MHz \citep{ocran2020}. \citet{chakraborty2019} observed it with the upgraded GMRT at 300-500 MHz. \citet{sabater2020} present a comparison between these previous radio observations and our observations of the field. In addition, the ELAIS-N1 field has been covered by several large radio surveys, including the Westerbork Northern Sky Survey (WENSS, 325\,MHz, \citealt{rengelink1997}), the NRAO Very Large Array  Sky Survey (NVSS, 1.4\,GHz, \citealt{condon1998}), and the Faint Images of the Radio Sky at Twenty Centimeters Survey (FIRST, 1.4\,GHz, \citealt{becker1995,white1997}).

Radio source counts are a useful tool for studying the evolution of radio source populations. They have been well studied in total intensity (e.g., \citealt{hopkins2003,condon2012,smolcic2017}), but polarized radio source counts are under-explored, in particular at low frequencies and at sub-mJy flux densities. Polarized source counts probe the evolution of magnetic fields and, at low flux densities, are thought to be sensitive to properties of the various radio source populations in a different way than the total-intensity source counts, since different populations are expected to show different intrinsic polarization properties related to the different mechanisms of emission \citep{stil2009,hardcastle1997,laing2014}. \citet{taylor2007} presented polarimetric observations of the ELAIS-N1 field at 1.4\,GHz. They imaged a region of 7.43\,deg$^{2}$ to a maximum sensitivity in Stokes $Q$ and $U$ of 78\,$\mu$Jy~beam$^{-1}$, and detected 83 polarized sources. \citet{grant2010} presented an extension of that study, imaging a region of 15.16\,deg$^{2}$, also at 1.4\,GHz, to a maximum sensitivity in Stokes $Q$ and $U$ of 45\,$\mu$Jy~beam$^{-1}$, and detected 136 polarized sources. They constructed the Euclidean-normalized polarized differential source counts down to 400~$\mu$Jy and found that fainter radio sources have a higher fractional polarization than the brighter ones. This result was questioned by \citet{hales2014,hales2014b}, who described the second data release of the Australia Telescope Large Area Survey (ATLAS) at 1.4~GHz and presented the 1.4~GHz Euclidean normalized differential number-counts from observations of the Chandra Deep Field-South (CDFS) and European Large Area Infrared Space Observatory Survey-South 1 (ELAIS-S1) regions. They found a smooth decline in both the total-intensity and linear polarization source counts from mJy levels down to $\sim$\!100~$\mu$Jy (their survey limit) and no signs of an anticorrelation between total-intensity and fractional polarization. 

The precision in RM measurements largely depends on the span in wavelength-squared, $\lambda^{2}$, and the sensitivity of the observations. Therefore, very-low-frequency observations allow us to measure Faraday depths very precisely (typical precision on the order of  $0.1\,\mathrm{rad\,m^{-2}}$; see \citealt{brentjens2005}). However, the degree of polarization of most sources decreases with decreasing frequency due to Faraday depolarization \citep{burn1966}. \citet{bernardi2013} presented a Stokes $I$, $Q$, and $U$ survey at 189~MHz with the Murchison Widefield Array \citep[MWA;][]{tingay2013} covering 2400~deg$^{2}$ and found that the polarization fraction of compact sources decreases at lower frequencies. \citet{lenc2016} presented deep polarimetric observations at 154~MHz with the MWA, covering 625\,deg$^{2}$, and detected four extragalactic polarized sources. A follow-up wide-area MWA survey effort based on the GaLactic and Extra-Galactic All-Sky MWA  (GLEAM) survey \citep{hurley-walker2017}, called the POlarization from the GLEAM Survey \citep[POGS;][]{riseley2018}, led to the detection of 81 sources in 6400~deg$^{2}$ at 216~MHz, and 517 sources over the full southern sky at 169$-$231~MHz \citep{riseley2020}. However, these studies have relatively low angular resolution ($\approx3-16$\,\,arcmin), which can make the detection of polarized sources at low frequencies even more challenging due to beam depolarization. 

In general, recent developments in low-frequency radio interferometers and improvements in calibration techniques have enabled the detection of many extragalactic polarized sources. In particular, LOFAR has the potential to minimize the effect of beam depolarization with observations at higher angular resolution. \citet{neld2018} developed a computationally efficient and rigorously defined algorithm to find linearly polarized sources in LOFAR data and applied it to previously calibrated data of the M51 field \citep{mulcahy2014}. \citet{vaneck2018} reported developments in polarization processing for the 570~deg$^{2}$ preliminary data release region from the LOFAR Two-Metre Sky Survey (LoTSS\footnote{\url{https://lofar-surveys.org}}, \citealt{shimwell2017,shimwell2019}) and presented a catalog of 92 polarized radio sources at 150~MHz with 4.3-arcmin resolution and 1~mJy root mean square (rms) sensitivity. \citet{osullivan2019} presented the Faraday RM and depolarization properties of a giant radio galaxy using LOFAR and demonstrated the potential of LOFAR to probe the weak signature of the intergalactic magnetic field (see also \citealt{osullivan2020}).

In order to study magnetic fields in extragalactic sources, it is essential to reach similar sensitivity to the surveys at higher frequencies, but with higher precision in RM. This requires very deep observations at low frequencies, well below the thermal noise of a typical eight-hour LOFAR observing run. Therefore, in order to improve the signal-to-noise ratio (S/N) of our data and detect fainter sources, we tested and implemented a new stacking procedure to combine the Stokes $Q$ and $U$ data cubes obtained from several LOFAR observations.

Here, we present the methods and results using LOFAR images of a 16~deg$^{2}$ area toward the ELAIS-N1 field. The structure of this paper is as follows. In Sect.~\ref{sec:cal}, we give an overview of the radio polarization observations and of the data calibration process. The description of the stacking technique is presented in Sect.~\ref{sec:stack}. In Sect.~\ref{sec:res}, we present our catalog and discuss the properties of the detected sources, and we summarize our findings in Sect.~\ref{sec:con}. Throughout this paper, we assume a flat Lambda cold dark matter $(\Lambda$CDM) cosmology with ${\it{H}}_{0}$\,=\,67.4\,km\,s$^{-1}$\,Mpc$^{-1}$, $\Omega_{M}$\,=\,0.315, and $\Omega_{\Lambda}$\,=\,0.685 (\citealt{planck2018}).

\section{Observations and data processing}
\label{sec:cal}

\subsection{Observations}

A polarimetric analysis of the ELAIS-N1 field with LOFAR was initially carried out as part of commissioning activities to characterize the LOFAR performance and the foregrounds in the LOFAR EoR fields \citep{jelic2014}. Since then, significantly more data have been accumulated. Here we present observations of the ELAIS-N1 field, as part of the LoTSS Deep Field project. The LoTSS Deep Fields comprise substantially deeper observations of the four best-studied degree-scale extragalactic fields in the northern sky, with an ultimate aim of reaching an rms depth of $\sim$\!10~$\mu$Jy~beam$^{-1}$ over 30-50 deg$^2$. The LoTSS Deep Fields first data release \citep{tasse2020,sabater2020} comprises 100-200 hours of data on each of three of these fields (ELAIS-N1, Lockman Hole and Boötes).

The observations of the ELAIS-N1 field used for this project were carried out with the LOFAR High Band Antenna (HBA) between May 2013 and August 2015 (cycles 0, 2, and 4; proposals  LC0\_019, LC2\_024, and LC4\_008). A total of 27 epochs, where an epoch is defined as an eight-hour LOFAR observation, were observed in full polarization. However, five data sets were excluded because of poor ionospheric conditions, noise levels, or data quality. This resulted in 22 available epochs and a total of 176\,hours of observations. The observations were followed by 5 to 10 minute observations of 3C380, which was the source selected for the primary calibration process. For all epochs, the observations were centered at RA\,=\,16$^{h}$11$^{m}$ and Dec\,=\,55\degree00$^{\prime}$. A detailed description of the observations can be found in \citet{sabater2020}.

In this paper, we only used  six epochs for testing purposes and to validate our new procedures on a representative subset of the available data. The Stokes $Q$ and $U$ data cubes each have 640 frequency channels with a channel width of 97.7~kHz, covering a frequency range from 114.9~MHz to 177.4~MHz. The imaged region covers an area of 16~deg$^{2}$. The identification numbers of the selected epochs used here are: 020, 024, 027, 028, 030, and 031. The other epochs either had a different frequency resolution or showed some artifacts in their polarized intensity images, which will require further analysis. We will present results from the combination of a larger set of observations in a follow-up paper.

\subsection{Data calibration}
\label{sec:calib}

New calibration pipelines and software were required in order to deal with the data size (up to 4~TB for each data set) and with the correction for direction-dependent ionospheric effects \citep{intema2009}. First, the data were calibrated using the software \texttt{PREFACTOR}\footnote{\url{https://github.com/lofar-astron/prefactor}}. This pipeline was developed to correct for various instrumental and ionospheric effects in LOFAR observations. It performs a direction-independent calibration and prepares the data to be used for any subsequent direction-dependent calibration. The procedure is described by \citet{degasperin2019}; see also \citet{vanweeren2016} and \citet{williams2016}. The calibrator source 3C380 was used. The AO\texttt{FLAGGER} from \citet{offringa2012} was used to flag the data for radio frequency interference (RFI). The  RM\texttt{EXTRACT}\footnote{\url{https://github.com/lofar-astron/RMextract}} software \citep{mevius2018} was used to correct the data for temporal variation in ionospheric Faraday rotation. \citet{sotomayor2013} estimated uncertainties of 0.1\,--\,0.3~rad~m$^{-2}$ after correcting for ionospheric Faraday rotation. The resulting calibration solutions were then applied to the target field. 

Once the direction-independent calibrated data were obtained, they were further processed to correct direction-dependent effects using the latest version of the DDF-pipeline\footnote{\url{https://github.com/mhardcastle/ddf-pipeline}} \citep{tasse2020}. This pipeline was developed to produce high-fidelity images for the LOFAR surveys. It performs several iterations of direction-dependent self-calibration on the data using kMS\footnote{\url{https://github.com/saopicc/killMS}}, which is a direction-dependent radio interferometric calibration package \citep{tasse2014a,tasse2014b,smirnov2015}, and DDF\texttt{ACET}\footnote{\url{https://github.com/saopicc/DDFacet}}, which is a facet-based radio imaging package that takes into account generic direction-dependent effects \citep{tasse2018}. Among the final products of the pipeline are $Q,U$ cubes generated for each epoch at an angular resolution of 20$^{\prime\prime}$. The data cubes are not deconvolved (i.e., all sources are convolved with the dirty beam). This explains the structures seen around bright sources in polarized intensity. A detailed description of the data calibration for this project can be found in \citet{sabater2020}. 

\begin{figure}
\includegraphics[width=\linewidth]{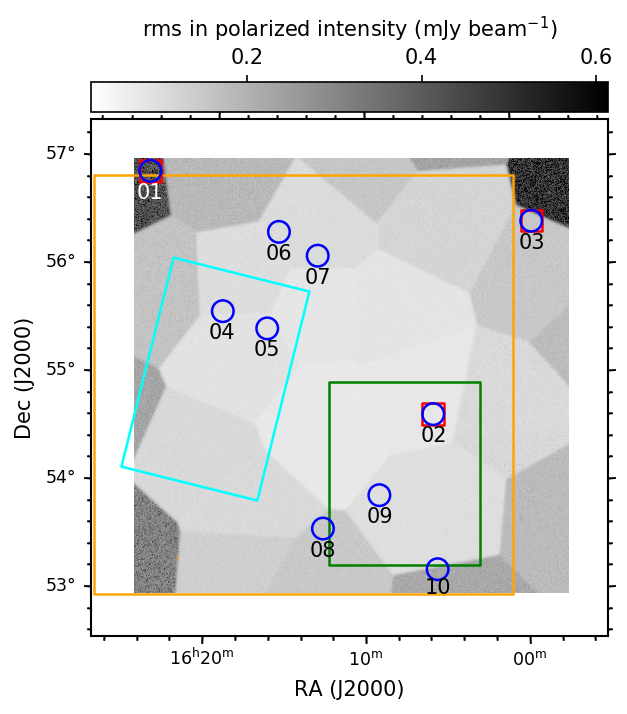}
\caption{Grayscale image of the noise in the polarized intensity map from the epoch 024 observations of ELAIS-N1 at 114.9-177.4 MHz. The pattern seen is due to the facet layout used in the DDF-pipeline. The angular resolution of the observations is 20$^{\prime\prime}$. The green square shows the region chosen to test our stacking technique (1.4~deg~$\times$~1.7~deg). The cyan square represents the ELAIS-N1 field originally defined by \citet{oliver2000}. The larger orange square represents the area imaged by \citet{grant2010} at 1.4~GHz. The red squares represent the locations of the detected polarized sources in the epoch 024 data alone. The blue circles represent the locations of the detected polarized sources after using the stacking technique for the six epochs described in the text.}
\label{fig:p_ref}
\end{figure}

In Fig.~\ref{fig:p_ref}, we show an image of the noise in the polarized intensity map ($\sigma_{\rm QU}$, see Sect.~\ref{sec:rmsyn}) of one of the epochs of the observations of the ELAIS-N1 field. We also outline the region imaged by \citet{grant2010} at 1.4~GHz (orange square), the region chosen to test our stacking technique (green rectangle, see Sect.~\ref{sec:stack}), and the location of the detected polarized sources (marked with red squares and blue circles, see Sect.~\ref{sec:cat}). The central coordinates of the LOFAR observations were offset from the region originally defined by \citet{oliver2000} as the ELAIS-N1 field (cyan rectangle) in order to better exploit multiwavelength surveys of the field \citep{kondapally2020}. In the noise map we can see the pattern created by the facet layout used in the DDF-pipeline \citep{tasse2020}, which solves and corrects the direction-dependent errors in a number of facets that cover the observed field of view. The facets appear clearly in the noise map because each calibration direction has independent calibration solutions and primary beam corrections. We expect higher rms noise with increasing distance from the pointing center because of the decrease in sensitivity associated with the primary beam response. This pattern is no longer visible using the latest version of DDF\texttt{ACET} (see e.g., \citealt{sabater2020}), where a continuous beam that gradually changes over the entire map is applied. However, an early version of the DDF\texttt{ACET} was used for the present data sets, where one beam correction per facet was applied rather than a continuous one. In a future publication, where it is intended to include all the epochs, the new processed data with the smooth beam will be used.

\subsection{RM synthesis}
\label{sec:rmsyn}

In the simplest scenario, in which the background radiation is Faraday rotated due to a foreground magneto-ionic medium, the Faraday depth is equivalent to the RM (see e.g., \citealt{mao2014}). Rotation measure synthesis \citep{brentjens2005} was performed on the Stokes $Q$ and $U$ data using \texttt{pyrmsynth\_lite}\footnote{\url{https://github.com/sabourke/pyrmsynth_lite}}, a modified version of \texttt{pyrmsynth}\footnote{\url{https://github.com/mrbell/pyrmsynth}} intended to analyze the polarization cubes produced as part of the second data release of LoTSS, developed by one of us (S. Bourke). Polarized intensity images and RM cubes with a span in Faraday depth, $\phi$, of $\pm$450\,rad\,m$^{-2}$ and a resolution of 0.3\,rad\,m$^{-2}$ were created. To avoid instrumental polarization, we excluded the range [$-3$,$+1.5$]\,rad\,m$^{-2}$. This range is asymmetric around zero\,rad\,m$^{-2}$ because of corrections for ionospheric Faraday rotation as described in Sect.~\ref{sec:calib}. \texttt{RMCLEAN} \citep{heald2009} was used after running the RM synthesis. Images in the Faraday depth range [350,450]\,rad\,m$^{-2}$, where no polarization signal is expected, were generated as a direct representation of the noise in our RM cubes, $\sigma_{\rm QU}$. We used these noise maps to calculate our sensitivity. 

The resolution in Faraday depth, $\delta\phi$, of our data is 0.9~rad~m$^{-2}$, the maximum observable Faraday depth, |$\phi_{\rm max}$|, is 300~rad\,m$^{-2}$, and the largest scale in $\phi$ space to which our data are sensitive is 1.07~rad\,m$^{-2}$. These numbers were calculated using Eqs.~61-63 from \citet{brentjens2005}. Due to bandwidth depolarization, we lose sensitivity to large RM sources (at the low-frequency end). However, we do not expect this to be an issue for this field because the expected RM contribution from the Galaxy is low (<~40~rad\,m$^{-2}$; e.g., \citealt{oppermann2015}).

\begin{figure}
\includegraphics[width=\linewidth]{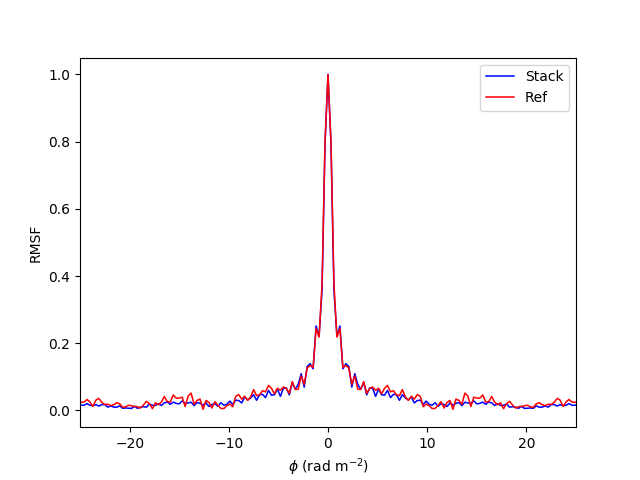}
\caption{Rotation measure spread function (RMSF) corresponding to the observations used in this paper. The data from the reference epoch alone are shown in red, and from the stacked data in blue (see text for more details).}
\label{fig:rmsf}
\end{figure}

Figure~\ref{fig:rmsf} shows the rotation measure spread function (RMSF) for the reference epoch (i.e., a single observation run) and the stacked data (of six observation runs; see Sect.~\ref{sec:stack} for details on the reference epoch and the stacking technique) of the ELAIS-N1 observations. The measured full width at half maximum (FWHM) of the RMSF is 0.9~rad\,m$^{-2}$ (confirming the computed resolution in Faraday depth), and the increment in Faraday depth is 0.3\,rad\,m$^{-2}$. The noise of the stacked data is lower than that of the reference epoch.

\subsection{Source extraction}
\label{sec:sext}

An in-depth analysis of the completeness and reliability of a comprehensive catalog of the polarized sources in the field is beyond the scope of this paper. This will be a matter for a follow-up paper. Instead we focus on the most reliably identified sources, using a detection threshold of 8$\sigma_{\rm QU}$. This threshold is based on the results of \citet{george2012}, who analyzed the false detection rates as a function of S/N and found the false detection rate to be less than 10$^{\rm -4}$ for an 8$\sigma_{\rm QU}$ detection threshold.

To identify the locations of polarized sources in the field, we divided the polarized intensity map by the noise map, both resulting from running RM synthesis, and derived a S/N map. We then searched the S/N map for pixels above the detection threshold. Each group of connected pixels (or island) was visually inspected in the polarized intensity map and identified in the Stokes $I$ map to ensure a reliable detection. If only one isolated pixel was found above the threshold and/or no Stokes $I$ counterpart was associated with an island, the detection was rejected. The Stokes $I$ maps are presented in \citet{sabater2020}.

\begin{figure}
\centering
\begin{subfigure}{0.35\textwidth}
    \includegraphics[width=\linewidth]{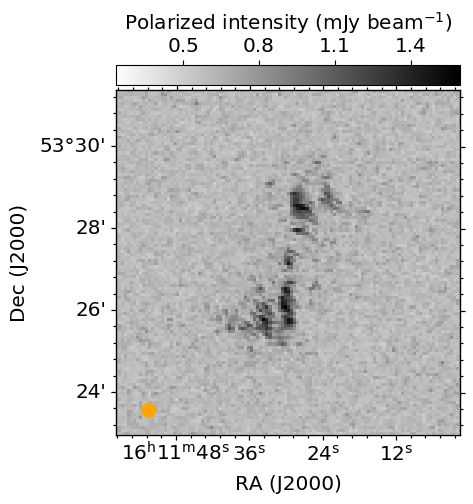}
\end{subfigure}
\begin{subfigure}{0.35\textwidth}
    \includegraphics[width=\linewidth]{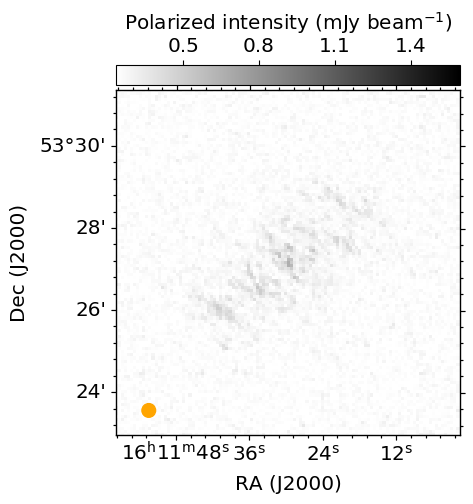}
\end{subfigure}
\caption{Polarized intensity maps of a source rejected as spuriously polarized by the criterion of $>$0.2\% fractional polarization, for the reference epoch (top) and stacked data (bottom).}
\label{fig:leak}
\end{figure}

We do not expect the instrumental polarization to be greater than 0.2\%. Therefore, we calculated the observed fractional polarization, $\Pi$ (defined as the ratio of the peak polarized intensity to the peak intensity in Stokes $I$, both at 20$^{\prime\prime}$), and used this threshold to reject spurious polarized sources. We used the Stokes $I$ map at 20$^{\prime\prime}$ angular resolution \citep{sabater2020} to calculate this fraction. Six sources were removed from the final catalog of detected sources since they had a fractional polarization lower than 0.2\%. One of these sources is shown as an example in Fig.~\ref{fig:leak}. For this source, we find that the polarized intensity decreases substantially after the stacking procedure described in Sect.~\ref{sec:stack}, consistent with the conclusion that it is artificially polarized. We present the catalog of detected polarized sources in Sect.~\ref{sec:cat}.

\section{Stacking technique}
\label{sec:stack}

Since we do not perform a full polarization calibration, possible changes in the polarization angle from different observing runs may be present. Taking into account that there is no absolute polarization angle calibration, we used a small region of the field (2.4~deg$^{2}$, see Fig.~\ref{fig:p_ref}) to test several approaches to the polarization angle alignment and stacking. This minimized the time needed to process the polarization data (to obviate the alternative of several weeks of computing time to process the full 16~deg$^{2}$). We chose this region because it contains the strongest polarized source in the field (with a peak polarized intensity, P$_{p}$, of $\sim$\!5\,mJy\,beam$^{-1}$) as well as a faint, potentially polarized source (with a S/N of $\sim$\!7 in a single epoch). 

The stacking technique presented here consists of the combination of the individual Stokes $Q$ and $U$ frequency channel images of each observed epoch. In other words, the data cube of each epoch is separated into the individual $Q$ and $U$ channel maps, and we stack each channel of each epoch together. For this, we used the toolkit for assembling Flexible Image Transport System (FITS) images into custom mosaics, \texttt{Montage}\footnote{\url{http://montage.ipac.caltech.edu}} \citep{berriman2003}. In our case, since all our epochs are centered at the same coordinates, we can use it to stack our data.

As noted above, due to ionospheric correction effects, the polarization angle may differ between separate epochs. Therefore, we need to align them before incoherently adding additional epoch data to avoid depolarizing the sources. To align the polarization angle, we used the 5\,mJy~beam$^{-1}$ polarized source as the reference source (later labeled as source 02, see Sect.~\ref{sec:cat}). We then calculated the reference polarized angle as the one corresponding to the coordinates of the peak pixel of the reference source in the polarized intensity image. These coordinates are RA\,=\,241.4080 and Dec\,=\,54.6551 (deg, J2000). The polarization angle, $\chi$, is calculated from the Stokes $Q$ and $U$:
\begin{eqnarray}
\chi = \frac{1}{2} \tan^{-1} \frac{U}{Q}
.\end{eqnarray}

\begin{figure}
\centering
\includegraphics[scale=0.6]{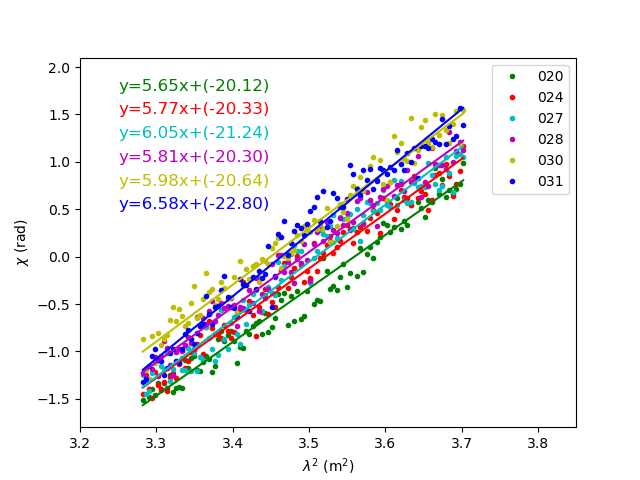}
\caption{Polarization angle, $\chi$, versus observing wavelength squared, $\lambda{^2}$, for a  subset of the full frequency range for each of the individual epochs, overlaid in different colors. The equations for the lines of best fit for each epoch are also shown.}
\label{fig:pa_ind}
\end{figure}

In order to choose the reference epoch, we plotted the polarization angle versus $\lambda{^2}$ for each epoch (see Fig.~\ref{fig:pa_ind}). With this plot we can see a roughly constant angle offset between the observations, as expected. This plot can also be used to flag or down-weight those epochs showing a slope or scatter outside the range of the others. We chose epoch 024 as the reference epoch, as it was the one with the smallest uncertainty in its gradient and the greatest S/N in its polarized intensity image. 

We attempted applying angle corrections using individual $Q$,$U$ values per channel or, alternatively, the average angle difference across a broad range of frequencies (by calculating the slope of the polarization angle as a function of $\lambda{^2}$ of one frequency subband), but these methods were not ideal. The main disadvantages are that the polarization angle correction is sensitive to channel-dependent noise and/or artifacts, and the shift in polarization angle depends on the frequency range chosen. Instead, we used a method based on the angles determined through the coherent addition of the signal across the full band using RM synthesis. However, future, more advanced correction techniques could also include a combination of several methods.

We calculated the polarization angles using the peak pixel of the reference source in the coherently band-averaged Stokes $Q$,$U$ images following RM synthesis. This provides the polarization angle at the average wavelength-squared value, $\lambda^{2}_{0}$. Since the number of flagged channels differs between the epochs, the value of $\lambda^{2}_{0}$ is slightly different for different epochs. Therefore, in order to calculate the difference between the reference polarization angle and the one of the epoch to be corrected, first we have to calculate the angle of the latter at the reference $\lambda^{2}_{0}$. For this purpose, we used the RM value of the RM map output from RM synthesis at the same peak pixel. In this case, we ran RM synthesis for each epoch, using the range of $\pm$10\,rad~m$^{-2}$ (the RM value is at $\sim$6\,rad~m$^{-2}$) with a sampling of 0.01\,rad~m$^{-2}$ (i.e., higher than described in Sect.~\ref{sec:rmsyn}). This was done to achieve better Faraday depth sampling for a more precise correction, as we expect the RMs to be different by 0.1 to 0.3 rad~m$^{-2}$ after the ionospheric Faraday rotation correction (e.g., \citealt{sotomayor2013}). Table~\ref{table:rm_lambda} shows the date at which the observation started \citep{sabater2020}, $\lambda^{2}_{0}$ values, and the RM measured at the peak pixel of the reference source and its uncertainty (calculated as the resolution in Faraday depth divided by twice the S/N of the detection \citep{brentjens2005}) for each epoch. 
 
\begin{table}
\caption{Summary of observation date, $\lambda^{2}_{0}$, RM of the reference source (using a higher sampling of Faraday depth, see text for details), and applied polarization position angle correction for each observed epoch.}             
\label{table:rm_lambda}      
%\centering                         
\begin{tabular}{c c c c c}      
\hline\hline               
Epoch & Date & $\lambda^{2}_{0}$  & RM  & $\Delta\chi_{\rm corr}$  \\ 
 & & [m$^{2}]$ & [rad~m$^{-2}$] & [deg] \\
\hline                       
  020 & 2015-06-07 & 4.412 & 5.86$\pm$0.03 & 17.8$\pm$1.7 \\
  024 & 2015-06-19 & 4.371 & 5.91$\pm$0.03 & `reference' \\
  027 & 2015-06-29 & 4.414 & 5.94$\pm$0.03 & -1.7$\pm$1.7 \\
  028 & 2015-07-01 & 4.413 & 5.95$\pm$0.02 & -10.3$\pm$1.8 \\
  030 & 2015-08-07 & 4.346 & 5.99$\pm$0.05 & -25.8$\pm$1.8 \\
  031 & 2015-08-22 & 4.413 & 6.03$\pm$0.04 & -15.5$\pm$1.8 \\
\hline                                 
\end{tabular}
\end{table}

We then calculated the polarization angle of the epoch to be corrected at the reference $\lambda^{2}_{0}$, $\chi_{\rm ep}(\lambda_{\rm 0,ref}^2)$ (where {\it{ep}} refers to the epoch to be corrected), as follows:
\begin{eqnarray}
\chi_{\rm ep}(\lambda_{\rm 0,ref}^2) = 
\chi_{\rm ep}(\lambda_{\rm 0,ep}^2) 
+ {\rm RM}_{\rm ep} \cdot (\lambda_{\rm 0,ref}^2 
- \lambda_{\rm 0,ep}^2)
,\end{eqnarray}
\noindent where $\chi_{\rm ep}(\lambda_{\rm 0,ep}^2)$ is the polarization angle of the epoch to be corrected, calculated from the resulting $Q$ and $U$ maps after running RM synthesis on that epoch, RM$_{\rm ep}$ is the RM taken from the resulting RM map from RM synthesis of the epoch to be corrected, $\lambda_{\rm 0,ref}^{2}$ is the average wavelength-squared value of the reference epoch, and $\lambda_{\rm 0,ep}^{2}$ is the average wavelength-squared value of the epoch to be corrected.

Once we calculated the reference polarization angle and the angle of the epoch to be corrected, we applied the difference, 
\begin{eqnarray}
\Delta\chi_{\rm corr,ep} = \chi_{\rm ref}(\lambda_{\rm 0,ref}^2) - \chi_{\rm ep}(\lambda_{\rm 0,ref}^2)
,\end{eqnarray}

\noindent to each pixel of each channel in Stokes $Q$ and $U$ using the following relations:
\begin{eqnarray}
Q_{\rm corr,ep} = p_{\rm ep} \cos 2 (\chi_{\rm ep} + \Delta\chi_{\rm corr,ep}) 
,\end{eqnarray}
\begin{eqnarray}
U_{\rm corr,ep} = p_{\rm ep} \sin 2 (\chi_{\rm ep} + \Delta\chi_{\rm corr,ep}) 
.\end{eqnarray}

\noindent Here, $Q_{\rm corr,ep}$ and $U_{\rm corr,ep}$ are the corrected Stokes $Q$ and $U$ channels and $p_{\rm ep}$ is the polarized intensity of the epoch to be corrected:
\begin{eqnarray}
p_{\rm ep} = \sqrt{Q^{2}_{\rm ep} + U^{2}_{\rm ep}}
.\end{eqnarray}

The polarization angle corrections as applied to each epoch using this method, and their errors, are listed in Table~\ref{table:rm_lambda}. The errors in the polarization angle correction have been calculated following error propagation rules, where the errors in $Q$ and $U$ are the rms values of the resulting $Q$ and $U$ maps from RM synthesis, respectively.

\begin{figure*}
\begin{subfigure}{0.5\textwidth}
    \includegraphics[width=.9\linewidth]{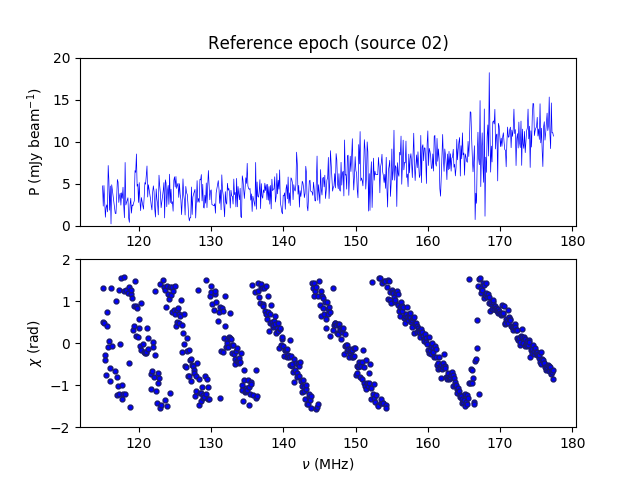}
\end{subfigure}
\begin{subfigure}{0.5\textwidth}
    \includegraphics[width=.9\linewidth]{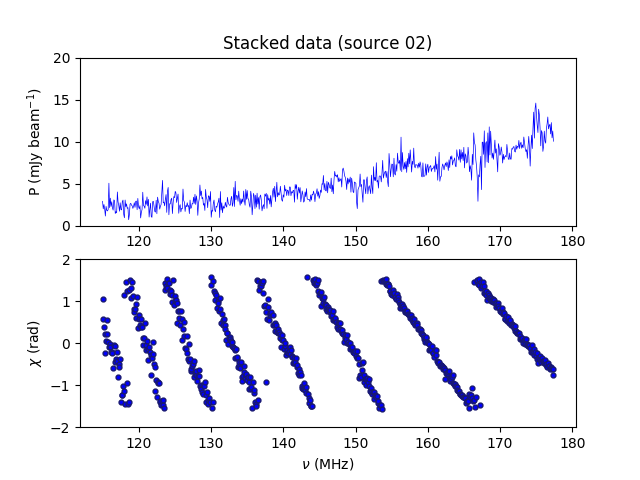}
\end{subfigure}
\begin{subfigure}{0.5\textwidth}
    \includegraphics[width=.9\linewidth]{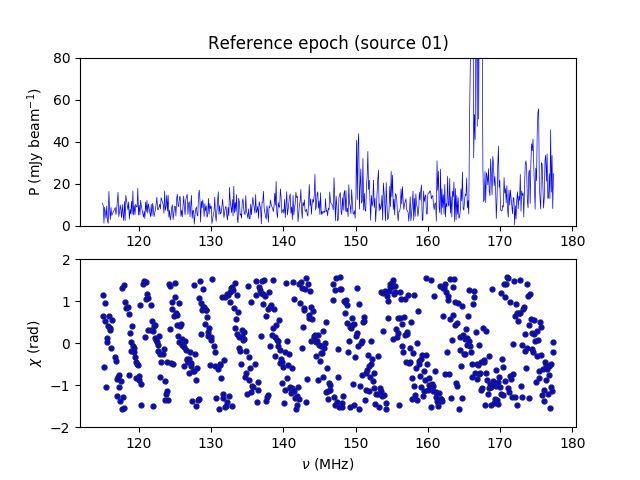}
\end{subfigure}
\begin{subfigure}{0.5\textwidth}
    \includegraphics[width=.9\linewidth]{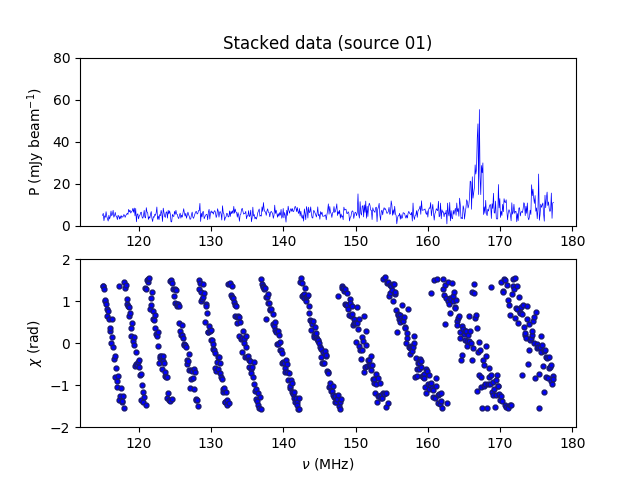}
\end{subfigure}
\caption{Polarized intensity and polarization angle versus frequency at the peak pixel of the reference source (named source 02) on the reference epoch (upper left) and the stacked data (upper right), and at the peak pixel of source 01 on the reference epoch (bottom left) and the stacked data (bottom right). The prominent spikes in the plots are likely due to RFI-related issues, and the different impact on the different sources might be due to the facet-based calibration.}
\label{fig:obst_rm}
\end{figure*}

Once the polarization angle corrections were applied to the data, we stacked the individual $Q$ and $U$ channels using \texttt{Montage} and ran RM synthesis on the stacked data with the parameters described in Sect.~\ref{sec:rmsyn}. Figure~\ref{fig:obst_rm} shows the polarized intensity and the polarization angle versus frequency at the peak pixel of the reference source on the reference epoch and the stacked data. The decrease in noise of the polarized intensity and the reduced scatter in the polarization angle versus wavelength squared after stacking the data are noticeable. In addition, to check if an angle correction using a single source as reference works well for the entire field, we also plot in Fig.~\ref{fig:obst_rm} the polarized intensity and the polarization angle versus frequency at the peak pixel of the source named later as 01 (see Sect.~\ref{sec:cat}) on the reference epoch and the stacked data. We chose this source because it is the farthest from the reference source (source 02, see Fig.~\ref{fig:p_ref}) with an angular separation of 3.46~deg. We can see the improvements on source 01 as well, demonstrating the validity of the stacking technique. We used this method to correct for the polarization angle on the entire field and stacked the $Q$,$U$ images for the six epochs used in this work.

\section{Results and discussion}
\label{sec:res}

\subsection{Stacked data versus reference epoch}

We computed the rms noise level (see Sect.~\ref{sec:rmsyn}) across the entire field (16~deg$^{2}$), and we obtained a minimum of 63\,$\mu$Jy~beam$^{-1}$ for the reference epoch, $\sigma_{\rm QU,ref}$, and a minimum of 26\,$\mu$Jy~beam$^{-1}$ for the stacked data (six epochs), $\sigma_{\rm QU,stack}$. The median rms noise level across the entire field is 111\,$\mu$Jy~beam$^{-1}$ for the reference epoch ($\overline {\sigma}_{\rm QU,ref}$) and 44\,$\mu$Jy~beam$^{-1}$ for the stacked data ($\overline {\sigma}_{\rm QU,stack}$). 
Therefore, the noise has been reduced by $\sim$\!$\sqrt{6}$ after stacking six epochs, as expected from Gaussian statistics. 

\begin{figure*}
\begin{subfigure}{0.5\textwidth}
    \centering
    \includegraphics[width=0.7\linewidth]{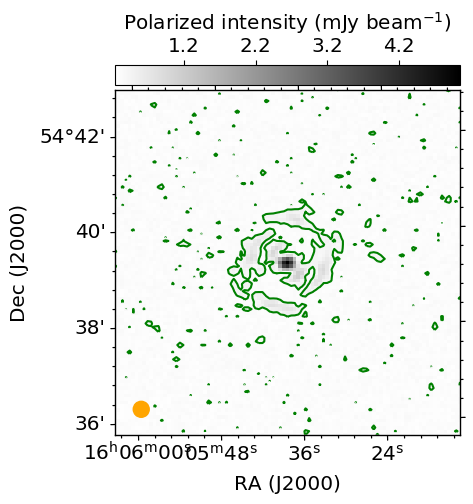}
\end{subfigure}
\begin{subfigure}{0.5\textwidth}
    \centering
    \includegraphics[width=0.7\linewidth]{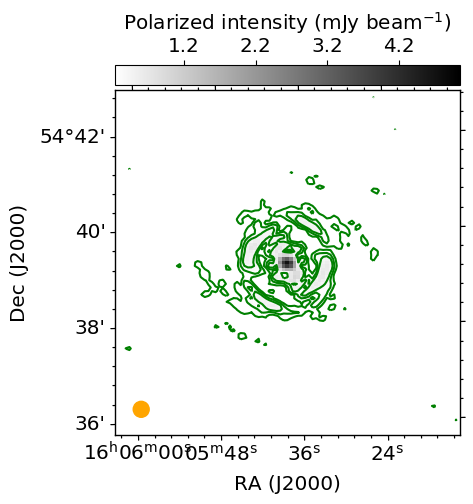}
\end{subfigure}
\begin{subfigure}{0.5\textwidth}
    \centering
    \includegraphics[width=0.7\linewidth]{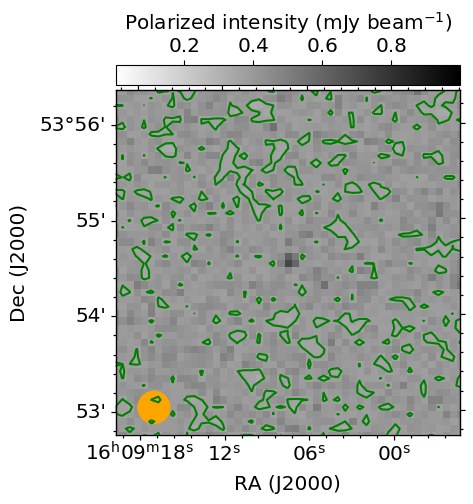}
\end{subfigure}
\begin{subfigure}{0.5\textwidth}
    \centering
    \includegraphics[width=0.7\linewidth]{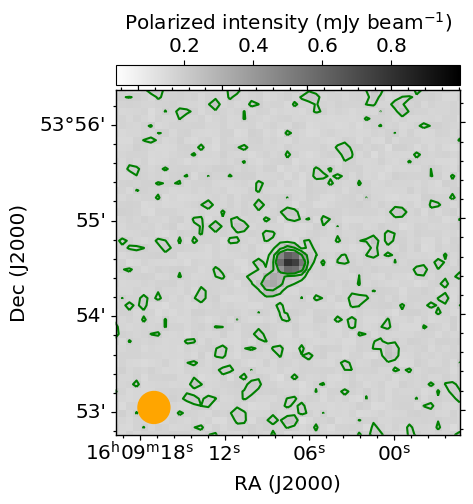}
\end{subfigure}
\caption{Polarized intensity map (not deconvolved) of the reference source (named 02, see text for details) (top panels) and of the source named 09 (bottom panels) before (left) and after (right) applying the stacking technique. Green contours showing the polarized intensity start at 3$\sigma_{\rm QU,ref}$ and increase by factors of $\sqrt{2}$. The orange circles in the bottom left corners represent the synthesized beam FWHM of the polarization observations (20$^{\prime\prime}$).}
\label{fig:obst_comp}
\end{figure*}

Figure~\ref{fig:obst_comp} shows the polarized intensity maps (not deconvolved) of the reference source and the source named later as 09 (see Sect.~\ref{sec:cat}) before and after applying the stacking technique, respectively. This further verifies the stacking technique because fainter sources are successfully detected; for example, the latter source that would not have been detected in a single epoch using a detection threshold of 8$\sigma_{\rm QU}$.

\subsection{Catalog of polarized sources}
\label{sec:cat}

Table~\ref{table:det_cat} lists the properties of the detected polarized sources in the ELAIS-N1 field. Using the source extraction method described in Sect.~\ref{sec:sext}, we detected ten polarized sources in our stacked data, seven of which were not detected in the reference epoch alone (using a detection threshold of 8$\sigma_{\rm QU}$). Therefore, with our resolution and sensitivity, we detect one polarized source per 5.3~deg$^{2}$ before stacking (in agreement with earlier estimates, e.g., \citealt{vaneck2018}) and one polarized source per 1.6~deg$^{2}$ after stacking our data. This value should be taken as a lower limit of the true low-frequency polarized source density at this sensitivity, considering that our catalog might be incomplete due to the high chosen detection threshold (see Sect.~\ref{sec:sext}) and the exclusion of the RM range near 0~rad\,m$^{-2}$. 

The peak polarized intensity of the source (columns (5) and (6) of our catalog, see Table~\ref{table:det_cat}) is corrected for polarization bias following \citet{george2012}. We used the rms at the peak pixel location of the source detection to calculate the S/N (columns (7) and (8) of our catalog). Columns (9) and (10) of our catalog show the RMs of the sources. The RM synthesis was run with the parameters described in Sect.~\ref{sec:rmsyn}. A parabola was fitted around the peak to improve the precision of the RM measurement. The error in the RM was calculated as the RM resolution divided by twice the S/N of the detection \citep{brentjens2005}. We note that the RM measurements presented in Tables~\ref{table:rm_lambda} and~\ref{table:det_cat} for the reference source 02 differ because of the different RM synthesis parameters used for the polarized angle correction (see Sect.~\ref{sec:stack}) and the default data processing (see Sect.~\ref{sec:rmsyn}).

\begin{table*}
\caption{Catalog of polarized sources in the ELAIS-N1 field detected with LOFAR at 146~MHz.}
\small
\label{table:det_cat}      
%\centering                         
\begin{tabular}{l l l l l l l l r r r l l}  
\hline\hline               
ID & ID$_{\rm StokesI}$ & RA$_{\rm pol}$ & Dec$_{\rm pol}$ & P$_{p,ref}$ & P$_{p,st}$ & S/N$_{ref}$ & S/N$_{st}$ & RM$_{ref}$ & RM$_{st}$  & S$_{p,I}$  & $\Pi_{ref}$ & $\Pi_{st}$ \\ 
 &  & [deg] & [deg]  &  [mJy/ & [mJy/  &  &  &  [rad m$^{-2}$] & [rad m$^{-2}$] &  [mJy/  & [\%] & [\%] \\
    &  & (J2000) & (J2000)  &  beam] & beam]  &  &  &   &  &  beam] &  &  \\
(1) & (2) & (3) & (4) & (5) & (6) & (7) & (8) & (9) & (10)  & (11)  & (12)  & (13) \\
\hline                       
01 & ILTJ162432.20  &  246.1327 & 56.8744  & 7.53 & 5.68 & 17 & 36 & 9.44$\pm$0.03 &  9.46$\pm$0.01  &  219.78 & 2.35 & 1.77 \\  
     & +565228.5 &  &   &  &  &  &  &   &  &   &  &   \\
02 & ILTJ160538.33  & 241.4080 & 54.6551  & 5.10 & 4.69 & 52 & 73 & 6.08$\pm$0.01 & 6.12$\pm$0.01  & 610.37  &  0.41 & 0.38  \\    
     & +543922.6  &  &   &  &  &  &  &   &  &   &  &   \\
03 & ILTJ155848.42  & 239.6998 & 56.4209  & 1.58 & 1.38 & 10 & 18 & --5.82$\pm$0.04 & --5.80$\pm$0.03 & 374.31  &  0.26 & 0.22 \\  
     & +562514.4  &  &   &  &  &  &  &   &  &   &  &   \\
04 & ILTJ161919.70  & 244.8342 & 55.6011  &      & 0.40 &    & 9  &     &  --4.70$\pm$0.05  &  60.72  &  & 0.33 \\     
     & +553556.7  &  &   &  &  &  &  &   &  &   &  &   \\
05 & ILTJ161623.79  & 244.0989 & 55.4513  &      & 0.50 &    & 14 &     &  --20.15$\pm$0.03  & 11.55  &  & 1.24 \\    
     & +552700.8  &  &   &  &  &  &  &   &  &   &  &   \\
06 & ILTJ161548.48  & 243.9340 & 56.3491  &      & 0.45 &    & 10 &     &  1.43$\pm$0.04*  &  32.97 &  & 0.55 \\     
     & +562029.8  &  &   &  &  &  &  &   &  &   &  &   \\
07 & ILTJ161314.05  & 243.2815 & 56.1325  &      & 0.40 &    & 9  &     &  --4.86$\pm$0.05  & 22.54  &  & 1.00 \\   
     & +560810.8  &  &   &  &  &  &  &   &  &   &  &   \\
08 & ILTJ161240.15  & 243.1669 & 53.5996  &      & 0.71 &    & 11 &     &  10.39$\pm$0.04 & 35.96  &  & 0.95 \\     
     & +533558.3  &  &   &  &  &  &  &   &  &   &  &   \\
09 & ILTJ160909.99  & 242.2811 & 53.9092  &      & 0.80 &    & 19 &     &  7.30$\pm$0.02  &  61.45  &  & 0.29 \\     
     & +535426.8  &  &   &  &  &  &  &   &  &   &  &   \\
10 & ILTJ160532.84 & 241.3855 & 53.2153 &      & 1.01 &    & 12 &     &  18.44$\pm$0.04  &  30.90 &  & 1.81 \\    
     & +531257.4   &  &   &  &  &  &  &   &  &   &  &   \\
\hline                                 
\end{tabular}
\begin{tablenotes}
\small
\item {\it {Col~1}}: Source name used in the present paper; {\it {Col~2}}: Source name used in the Stokes $I$ map, taken from \citet{sabater2020}; {\it {Cols~3,~4}}: Right ascension and declination (J2000) of the polarized source, in degrees; {\it {Cols~5,~6}}: Peak polarized intensity of the source, in mJy~beam$^{-1}$, in the reference epoch and the stacked data, respectively; {\it {Cols~7,~8}}: S/N of the source in the reference epoch and the stacked data, respectively; {\it {Cols~9,~10}}: Faraday RM of the source and its error, in rad~m$^{-2}$, in the reference epoch and the stacked data, respectively (RM synthesis parameters described in Sect.~\ref{sec:rmsyn}); {\it {Col~11}}: Peak intensity of the source on the Stokes $I$ map, taken from \citet{sabater2020} (6$^{\prime\prime}$ angular resolution), in mJy~beam$^{-1}$; {\it {Cols~12,~13}}: Fractional polarization of the source, calculated using both the polarization and the Stokes $I$ maps at 20$^{\prime\prime}$ angular resolution, on the reference epoch and the stacked data, respectively.
\item * The RM of source 06 had to be calculated separately since the peak of the Faraday spectrum is located very close to the limit of the instrumental polarization range.
\end{tablenotes}
\end{table*}

As mentioned before, polarized radio source counts are important for studying the changes in the properties of radio source populations. However, the number of detected sources presented in this paper is very low, and therefore this analysis will be discussed in a follow-up paper based on deeper images.

\begin{figure*}
\begin{subfigure}{0.22\textwidth}
    \includegraphics[width=\linewidth]{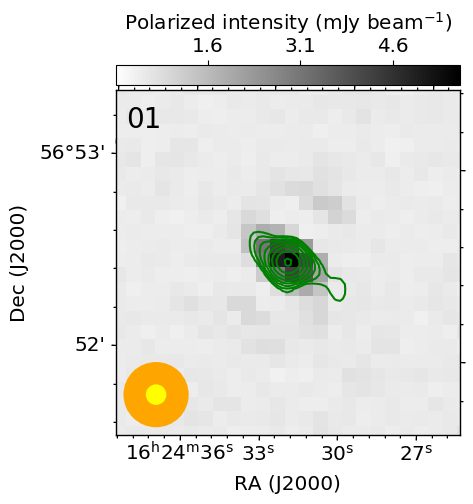}
\end{subfigure}
\begin{subfigure}{0.21\textwidth}
    \includegraphics[width=\linewidth]{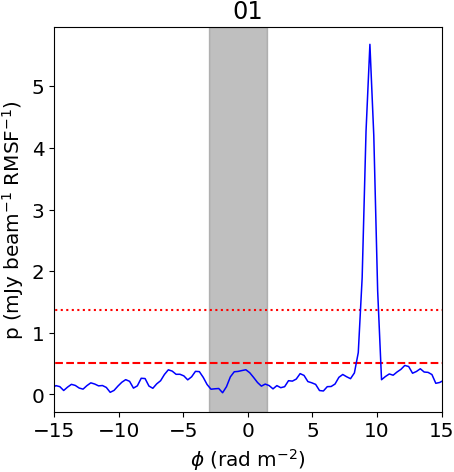}
\end{subfigure}
\begin{subfigure}{0.22\textwidth}
    \includegraphics[width=\linewidth]{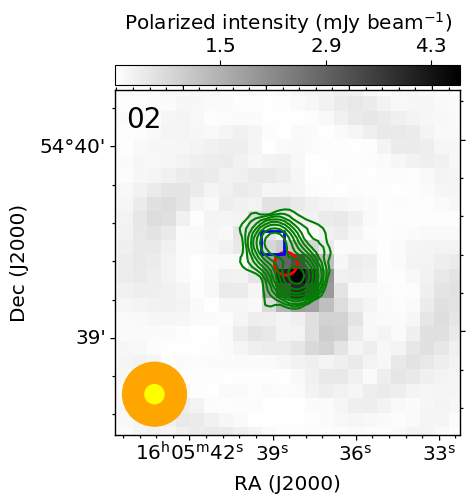}
\end{subfigure}
\begin{subfigure}{0.21\textwidth}
    \includegraphics[width=\linewidth]{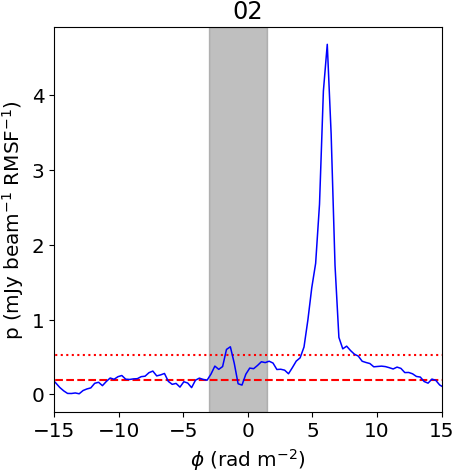}
\end{subfigure}
\begin{subfigure}{0.22\textwidth}
    \includegraphics[width=\linewidth]{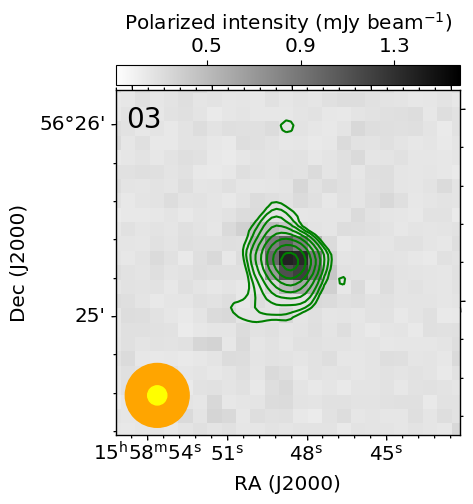}
\end{subfigure}
\begin{subfigure}{0.21\textwidth}
    \includegraphics[width=\linewidth]{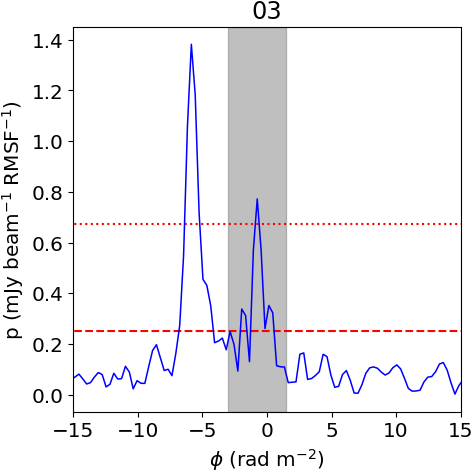}
\end{subfigure}
\begin{subfigure}{0.22\textwidth}
    \includegraphics[width=\linewidth]{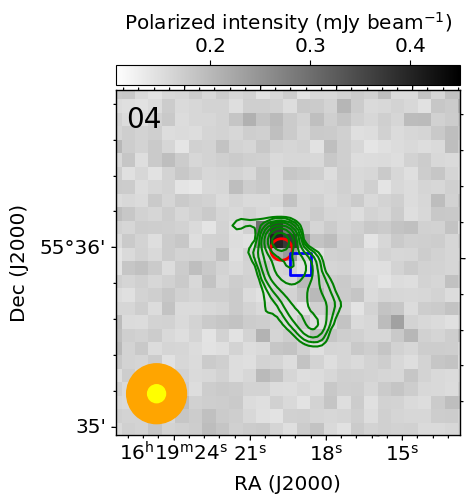}
\end{subfigure}
\begin{subfigure}{0.21\textwidth}
    \includegraphics[width=\linewidth]{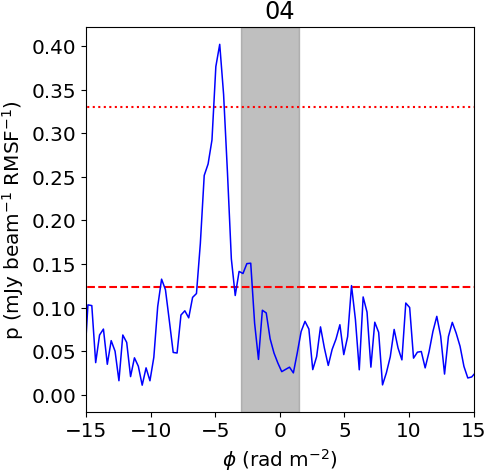}
\end{subfigure}
\begin{subfigure}{0.225\textwidth}
    \includegraphics[width=\linewidth]{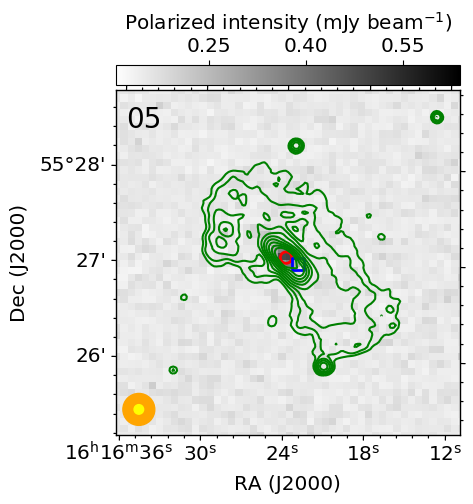}
\end{subfigure}
\begin{subfigure}{0.215\textwidth}
    \includegraphics[width=\linewidth]{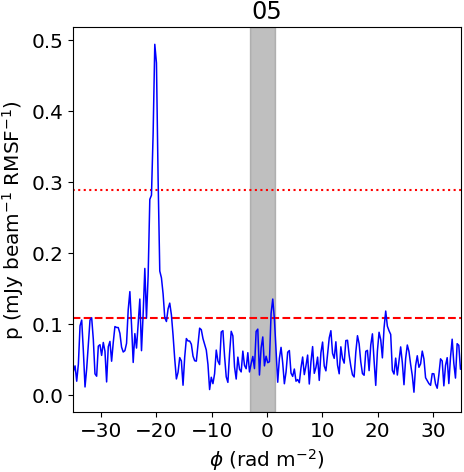}
\end{subfigure}
\begin{subfigure}{0.225\textwidth}
    \includegraphics[width=\linewidth]{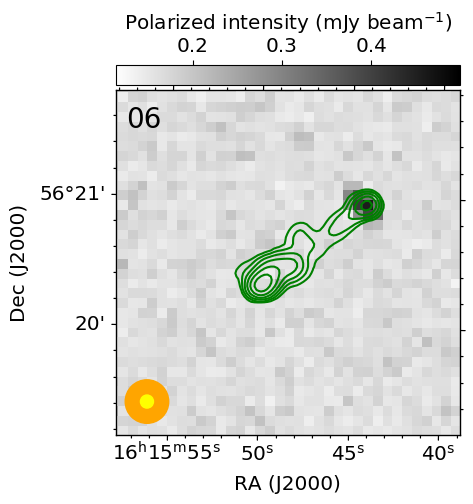}
\end{subfigure}
\begin{subfigure}{0.215\textwidth}
    \includegraphics[width=\linewidth]{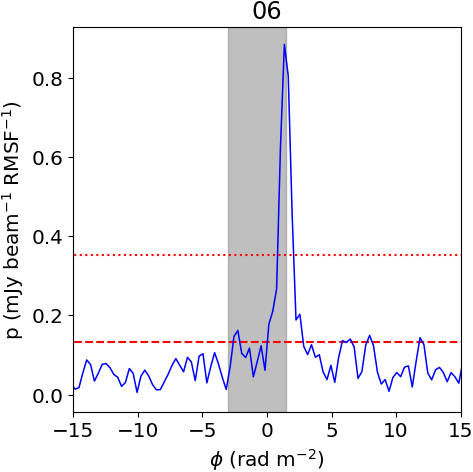}
\end{subfigure}
\begin{subfigure}{0.225\textwidth}
    \includegraphics[width=\linewidth]{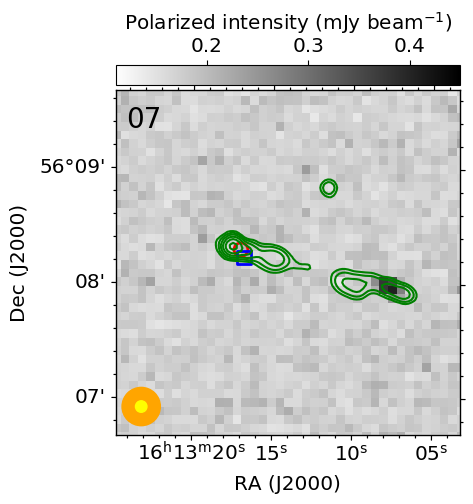}
\end{subfigure}
\begin{subfigure}{0.215\textwidth}
    \includegraphics[width=\linewidth]{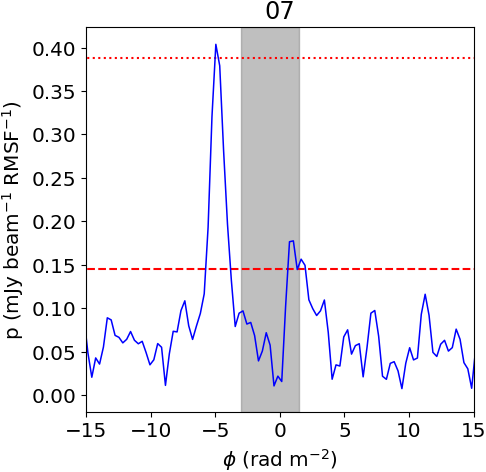}
\end{subfigure}
\begin{subfigure}{0.225\textwidth}
    \includegraphics[width=\linewidth]{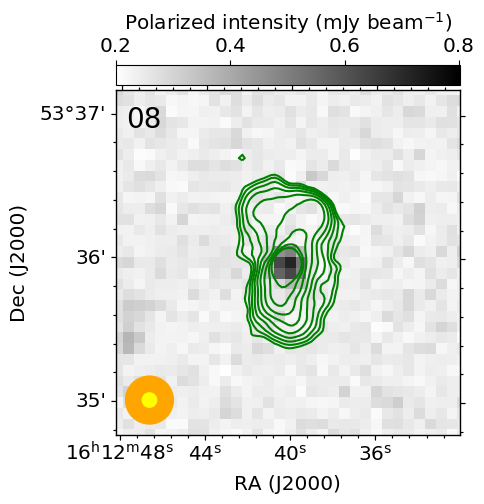}
\end{subfigure}
\begin{subfigure}{0.215\textwidth}
    \includegraphics[width=\linewidth]{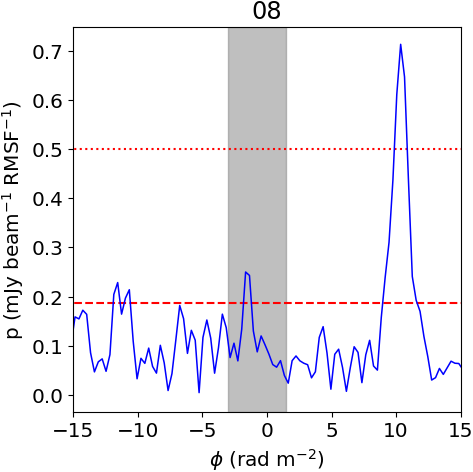}
\end{subfigure}
\begin{subfigure}{0.225\textwidth}
    \includegraphics[width=\linewidth]{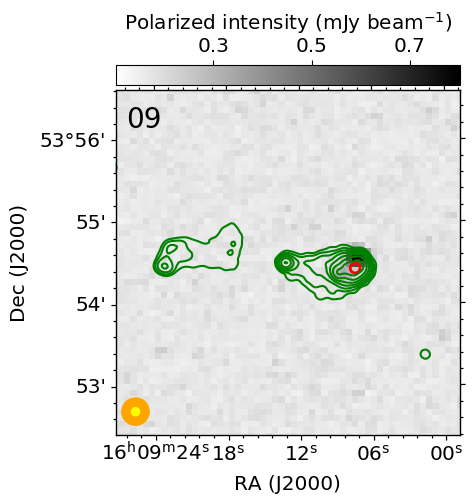}
\end{subfigure}
\hspace{0.5 cm}
\begin{subfigure}{0.215\textwidth}
    \includegraphics[width=\linewidth]{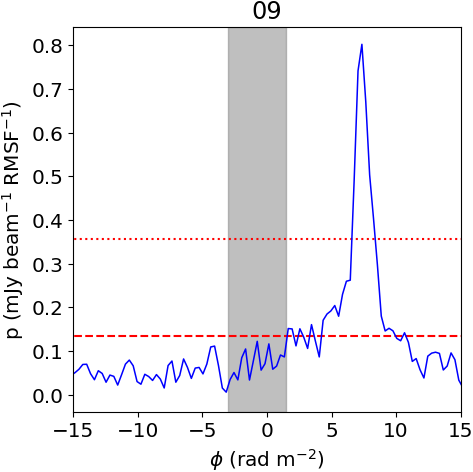}
\end{subfigure}
\hspace{0.5 cm}
\begin{subfigure}{0.225\textwidth}
    \includegraphics[width=\linewidth]{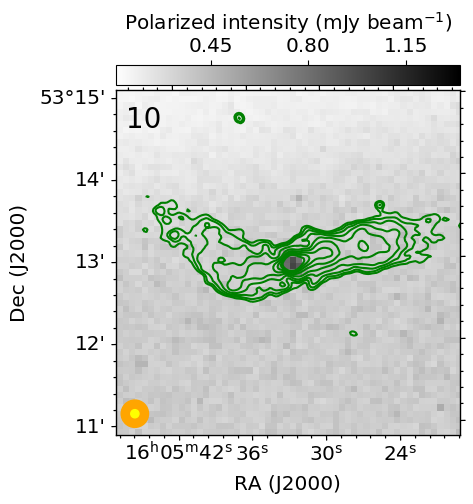}
\end{subfigure}
\hspace{0.5 cm}
\begin{subfigure}{0.215\textwidth}
    \includegraphics[width=\linewidth]{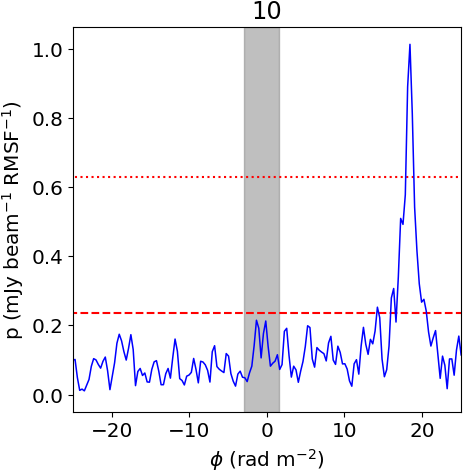}
\end{subfigure}
\caption{Polarized intensity maps with overlaid Stokes $I$ contours (green) and Faraday spectra of the LOFAR detected polarized sources. The orange and yellow circles in the bottom left corners represent the synthesized beam of the observations in polarization (20$^{\prime\prime}$) and in Stokes $I$ (6$^{\prime\prime}$), respectively. The Stokes $I$ contours start either at 55$\sigma_{\rm I}$ or 20$\sigma_{\rm I}$ and increase by factors of two, or at 10$\sigma_{\rm I}$ and increase by factors of $\sqrt{2}$ (see text for details). The blue square and red circle plotted on some of the polarized intensity maps represent the locations of the polarized sources detected at 1.4~GHz by \citet{taylor2007} and \citet{grant2010},  respectively. The dashed and dotted red horizontal lines plotted on the Faraday spectra represent the 3$\sigma_{\rm QU}$ and 8$\sigma_{\rm QU}$ levels, respectively. The vertical gray area encloses the excluded area in Faraday depth to avoid instrumental polarization leakage, sometimes visible as lesser peaks within this range.}
\label{fig:cutouts}
\end{figure*}

The LOFAR images of the detected polarized sources and their Faraday spectra at the location of the peak polarized intensity, using the stacked data, are shown in Fig.~\ref{fig:cutouts}. The Faraday spectra are zoomed in on to show the Faraday range containing the peak (only one Faraday depth component was found in all cases). We plotted the polarized intensity maps with the Stokes $I$ contours overlaid, starting at 55 times the median rms noise level of the Stokes $I$ map, $\sigma_{\rm I}$, and increasing by factors of two for sources 01, 02, and 03 (to avoid showing artifacts); at 20$\sigma_{\rm I}$ and increasing by factors of two for sources 04, 06, 07, and 09; and at 10$\sigma_{\rm I}$ and increasing by factors of $\sqrt{2}$ for the remaining sources. The angular resolution of the Stokes $I$ maps is 6$^{\prime\prime}$, and 20$^{\prime\prime}$ for the polarized intensity images. In addition, we overlaid horizontal lines on the Faraday spectra representing the 3$\sigma_{\rm QU}$ and 8$\sigma_{\rm QU}$ levels, using the derived $\sigma_{\rm QU}$ value at the location of the peak polarized intensity. We also highlight the area in Faraday depth excluded due to the instrumental polarization in gray. The locations of the detected sources in the sky are also shown in Fig.~\ref{fig:p_ref}.

Table~\ref{table:redsh} shows the redshifts for nine of the ten detected polarized sources. \citet{kondapally2020} cross-matched the PyBDSF\footnote{Python Blob Detection and Source Finder \citep{mohan2015}; \url{https://www.astron.nl/citt/pybdsf}} radio catalog from total-intensity observations of \citet{sabater2020} using various optical to infrared surveys. The LOFAR cross-identifications are used to then generate photometric redshifts and obtain existing spectroscopic redshifts as described by \citet{duncan2020}. Sources 01, 03, and 10 were outside of the multiwavelength surveys used for the radio cross-matching in \citet{kondapally2020}. Therefore, their counterparts were searched for using the NASA/IPAC Extragalactic Database (NED\footnote{\url{https://ned.ipac.caltech.edu/}}). The redshift of the counterpart associated with source 04 was flagged as possible, but doubtful, by \citet{kondapally2020}. For this reason, NED was used to estimate the redshift of source 04 as well. Nevertheless, the host ID for this source should be considered as uncertain. The corresponding references found in NED are shown in Table~\ref{table:redsh}. No counterpart for source 07 was found.

\begin{table}
%\centering 
\caption{Redshifts of the LOFAR-detected polarized sources.}
\small
\label{table:redsh}      
\begin{tabular}{l l l l l}  
\hline\hline               
ID  & ID$_{\rm optical}$ & $z$ & Type & Ref. \\
   &   &  \\
\hline                       
01  & SBS 1623+569  &   0.415  &  Ph & R09 \\        
02  & HELP J160538.300+543923.239  &  0.7911  &  Ph & D20 \\       
03  & WISE J155848.29+562514.2  &  0.300    &  Ph & F98 \\      
04*  & 87GB 161814.7+554307  &   0.405   &  Ph & R09 \\      
05  & SDSS J161623.52+552703.7  &   0.2430  &  Sp & D20 \\      
06  & WISEA J161547.94+562031.2  &   0.3347  &  Sp & D20 \\
07  & - & - & - & - \\
08  & WISEA J161240.16+533557.8  &  0.1346  &  Sp & D20 \\      
09  & WISEA J160913.19+535429.8  &  0.9928   &  Sp &  D20 \\      
10 & WISEA J160532.59+531257.6   &  0.0633  &  Sp & A17 \\      
\hline                                 
\end{tabular}
\begin{tablenotes}
\small
\item Ph = Photometric redshift
\item Sp = Spectroscopic redshift
\item References: F98 -- \citet{falco1998}; R09 -- \citet{richards2009}; A17 -- \citet{albareti2017}; D20 -- \citet{duncan2020}
\item *Optical host ID uncertain
\end{tablenotes}
\end{table}

The redshifts of the polarized sources provide the prospect of converting the measured RMs to that in the sources' rest frames (e.g., \citealt{michilli2018}). This requires the RM contribution from the Galactic foreground to be subtracted prior to the redshift correction. Since the uncertainties in the Galactic RM (GRM) foreground are comparable to the precise RM measurements in this work (see Sect.~\ref{sec:nvss}, Table~\ref{table:nvss}), we do not make this correction but note that this can be addressed in future work. Furthermore, this motivates the requirement for a denser and more accurate ``RM Grid'' to reconstruct the GRM foreground (e.g., \citealt{heald2020}).

\subsection{Analysis of results from the individual epochs}

\begin{figure}
\includegraphics[width=\linewidth]{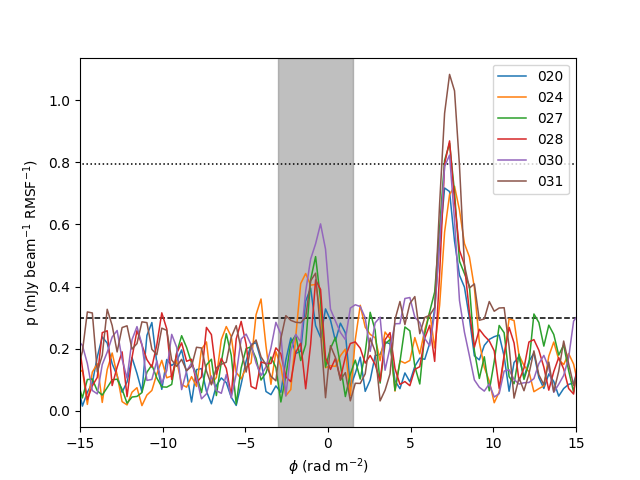}
\caption{Faraday spectra of source 09 for each epoch. The dashed and dotted black horizontal lines represent the 3$\sigma_{\rm QU}$ and 8$\sigma_{\rm QU}$ levels, respectively (see text for details). The vertical gray area represents the range around $\phi$=0~rad~m$^{-2}$ excluded from the analysis due to instrumental polarization.}
\label{fig:rm_ind}
\end{figure}

Here, we analyze the Faraday spectra and the characteristics of detected polarized sources for each individual epoch. Figure~\ref{fig:rm_ind} shows the Faraday spectrum of source 09 plotted for each observed epoch. The overlaid horizontal lines represent the 3$\sigma_{\rm QU}$ and 8$\sigma_{\rm QU}$ levels, using the median of the rms noise levels for each epoch. We can see that source 09 only satisfies the 8$\sigma_{\rm QU}$ detection threshold  for some of the epochs, demonstrating that the stacking technique is important for detecting those sources that are very close to the detection threshold limit. 

Table~\ref{table:ind} lists the polarized intensity, P$_{p}$, the RM, and the S/N (as described in Sect.~\ref{sec:cat}) for sources 02 and 09 measured at each epoch used here. The polarized intensity is given even if the S/N is below 8 (our detection threshold). It is also worth noting that the S/N does not behave in the same way for each epoch. As shown in Table~\ref{table:ind}, the epoch with the highest S/Ns for source 02 is not the same as the one for source 09.

\begin{table}
\caption{Measurements for sources 02 and 09 in each epoch. The RM synthesis parameters are described in Sect.~\ref{sec:rmsyn}.}
\label{table:ind}  
\small
%\centering                         
\begin{tabular}{l l l l l}      
\hline\hline               
ID & Epoch & P$_{p}$   & RM           & S/N  \\ 
  &  & [mJy beam$^{-1}$]  & [rad m$^{-2}$]  &      \\
\hline                       
 02  &   020   &  4.41  &   6.01$\pm$0.01  &   46.7  \\
 02  &   024   &  5.10  &   6.08$\pm$0.01  &   52.4 \\
 02  &   027   &  4.78  &   6.10$\pm$0.01  &   45.7 \\
 02   &  028   &  4.91  &   6.12$\pm$0.01  &   45.2 \\
 02  &   030   &  5.29   &  6.15$\pm$0.01  &   44.1 \\
 02  &   031   &  4.58  &   6.19$\pm$0.01  &   41.3 \\
 \hline
 09  &   020  &   0.73  &   7.18$\pm$0.06   &  7.9 \\
 09  &   024   &  0.71   &  7.58$\pm$0.06  &   7.1 \\
 09  &   027   &  0.85   &  7.29$\pm$0.05  &   8.5 \\
 09  &   028   &  0.86  &   7.28$\pm$0.06  &   7.1 \\
 09  &   030   &  0.82  &   7.25$\pm$0.06  &   7.2 \\
 09   &  031  &   1.08   &  7.41$\pm$0.04  &   10.3 \\
\hline                                 
\end{tabular}
\end{table}

The average polarized intensity for sources 02 and 09 over the six observing epochs is 4.84 and 0.84~mJy~beam$^{-1}$, respectively. However, the polarized intensity measured using the stacked data is 4.69~mJy~beam$^{-1}$ for source 02 and 0.80~mJy~beam$^{-1}$ for source 09. Therefore, the stacking technique seems to cause a small depolarization, which decreases the polarized intensity by $\sim$\!4\%. On the other hand, the decrease in polarized intensity is very small compared to the decrease in noise ($\sim$\!2.5 times lower), making the stacking technique a valuable tool to help detect fainter sources. Another source of the depolarization may be our inability to fully correct for the ionospheric Faraday rotation behavior over the observations. This approximate level of depolarization could be expected from residual RM errors (e.g., using Eq. (34) from \citealt{sokoloff1998}). 

\subsection{Measurements as a function of number of stacked epochs}

\begin{figure}
\includegraphics[width=\linewidth]{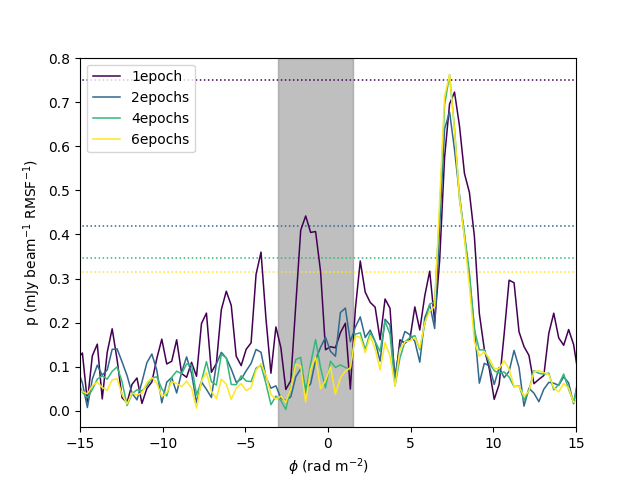}
\caption{Faraday spectra of source 09 for different numbers of stacked epochs. The dotted horizontal lines represent the corresponding 8$\sigma_{\rm QU}$ levels. The vertical gray area represents the range around $\phi$=0~rad~m$^{-2}$ excluded from the analysis due to instrumental polarization.}
\label{fig:rm_ev}
\end{figure}

Here, we analyze the effect of stacking a different number of epochs. We show source 09 as an example. Figure~\ref{fig:rm_ev} and Table~\ref{table:ev} show the evolution of the Faraday spectra by stacking a different number of epochs. The overlaid horizontal lines in Fig.~\ref{fig:rm_ev} represent the 8$\sigma_{\rm QU}$ levels, using the median of the rms noise level of the corresponding number of stacked epochs. In Table~\ref{table:ev}, the column N$_{\rm epoch}$ corresponds to the number of epochs stacked. We list the polarized intensity also in the case where the S/N is below 8 (our detection threshold).

\begin{table}
\caption{Measurements for source 09 using different numbers of stacked epochs. The RM synthesis parameters are described in Sect.~\ref{sec:rmsyn}.}
\label{table:ev}  
\small
%\centering                         
\begin{tabular}{l l l l l}      
\hline\hline               
ID & N$_{\rm epoch}$ & P$_{p}$   & RM           & S/N  \\ 
   &       & [mJy~beam$^{-1}$]  & [rad~m$^{-2}$]  &      \\
\hline                       
 09  &   1  &   0.71  &   7.58$\pm$0.06   & 7.1  \\
 09  &   2   &  0.68   &  7.27$\pm$0.03  &  13.2  \\
 09  &   4   &  0.76   &  7.31$\pm$0.03  &  17.1  \\
 09  &   6   &  0.80  &   7.30$\pm$0.02  &  18.7  \\
\hline                                 
\end{tabular}
\end{table}

The S/N of the source increases continuously with the number of epochs stacked. A large improvement is seen when comparing one epoch to the stacking of two epochs (the median noise goes down by more than $\sqrt{2}$), which we believe is due to the presence of the instrumental polarization close to $\phi=$~0. The instrumental polarization responses are uncorrelated between epochs, resulting in a larger than expected decrease in the noise.

\subsection{Comparison with 1.4~GHz measurements}
\label{sec:nvss}

We cross-matched our catalog of detected polarized sources with the catalog of \citet{grant2010}, who found 136 polarized sources at 1.4~GHz in the ELAIS-N1 field. Their observations had an angular resolution of 42$^{\prime\prime}$~$\times$~62$^{\prime\prime}$ (at the field center). We used a matching radius of 1.5$^{\prime}$ (to include sources similar to our source 07) and found five matches. In Table~\ref{table:comp}, we give the name, the peak intensity in Stokes $I$ at 146~MHz, S$_{p,146}$, the peak intensity in Stokes $I$ at 1.4~GHz, S$_{p,1420}$, the spectral index, $\alpha_{\rm 1420-325}$\footnote{$S(\nu)\propto \nu^{\alpha}$}, the fractional polarization at 146~MHz, $\Pi_{146}$, and the fractional polarization at 1.4~GHz, $\Pi_{1420}$, of these five sources. 

\begin{table}
\caption{146~MHz and 1.4~GHz properties of the detected polarized sources.}
\label{table:comp}  
\small
%\centering                         
\begin{tabular}{l l l l l l}      
\hline\hline               
ID & S$_{p,146}$ & S$_{p,1420}$ & $\alpha_{\rm 1420-325}$ &  $\Pi_{146}$ & $\Pi_{1420}$ \\ 
   & [mJy beam$^{-1}$]  & [mJy beam$^{-1}$]  &  &  [\%] & [\%] \\
\hline                       
02 & 610.37 & 218.97 & -0.89 & 0.38 & 4.68 \\
04 & 60.72  & 55.89  & -0.41 & 0.33 & 3.76 \\
05 & 11.55  & 12.10  & -0.95 & 1.24 & 8.03 \\
07 & 22.54  & 5.23   & -0.84 & 1.00 & 14.86 \\
09 & 61.45  & 43.84  & -1.01 & 0.29 & 1.47 \\
\hline                                 
\end{tabular}
\begin{tablenotes}
\small
\item The angular resolution of the 146~MHz observations is 20$^{\prime\prime}$ in polarization and 6$^{\prime\prime}$ in Stokes I. The angular resolution of the 1.4~GHz observations is 42$^{\prime\prime}$~$\times$~62$^{\prime\prime}$ (at the field center).
\item S$_{p,1420}$, $\alpha_{\rm 1420-325}$ and $\Pi_{1420}$ are taken from \citet{grant2010}. S$_{p,146}$ is taken from \citet{sabater2020}.
\item The fractional polarization at 146~MHz, $\Pi_{146}$, is calculated using both the polarization and the Stokes $I$ maps at 20$^{\prime\prime}$ angular resolution.
\end{tablenotes}
\end{table}

Figure~\ref{fig:p_ref} shows the overlapping area between our observations and the field imaged by \citet{grant2010} at 1.4~GHz. Sources 01 and 03 are outside of the area covered by \citet{grant2010}. Therefore, we have 1.4~GHz measurements for five out of eight polarized sources.

In Fig.~\ref{fig:cutouts}, we plot the locations of the polarized sources detected at 1.4~GHz by \citet{taylor2007} and \citet{grant2010} on our polarized intensity maps (when corresponding). We can see that the centroids from \citet{grant2010}, whose observations were more sensitive than the ones of \citet{taylor2007}, are more consistent with the LOFAR locations.

\citet{taylor2009} derived RMs for several tens of thousands of polarized radio sources using NVSS data. We cross-matched our catalog with theirs using a matching radius of 1.5$^{\prime}$ and found two counterparts. Table~\ref{table:nvss} shows the RM comparison between our data and the data from \citet{taylor2009}. The RM of source 01 is different at $\sim$4$\sigma$. However, since this source is a blazar, variability might play a role (see e.g., \citealt{anderson2019}). The RM of source 02 is consistent within $\sim$1$\sigma$. The GRM values at the location of each source are also shown in Table~\ref{table:nvss}. We obtained these values using the reconstruction of the Galactic Faraday depth sky from \citet{hutschenreuter2020}\footnote{\url{http://cdsarc.u-strasbg.fr/viz-bin/cat/J/A+A/633/A150}}. Eight of the LOFAR sources are consistent with each GRM value within $\sim$1$\sigma$, indicating that the Milky Way RM dominates the mean RM (as expected). Sources 05 and 10 are the most different ($\sim$4$\sigma$ and $\sim$2.4$\sigma$, respectively).

\begin{table}
\caption{RM comparison.}
\label{table:nvss}  
%\small
\centering                         
\begin{tabular}{r r r r}      
\hline\hline               
ID & RM & RM$_{\rm NVSS}$ & GRM \\ 
   & [rad m$^{-2}$]  & [rad m$^{-2}$] & [rad m$^{-2}$] \\
\hline 
01 &    9.46$\pm$0.01  & --7.1$\pm$4.2 &   1.79$\pm$5.64 \\                  
02 &    6.12$\pm$0.01  & --0.9$\pm$7.5 &   1.45$\pm$5.60 \\
03 &  --5.80$\pm$0.03  &               & --2.10$\pm$5.71 \\
04 &  --4.70$\pm$0.05  &               &   2.51$\pm$5.64 \\
05 & --20.15$\pm$0.03  &               &   2.05$\pm$5.46 \\
06 &    1.43$\pm$0.04  &               & --0.98$\pm$5.63 \\
07 &  --4.86$\pm$0.05  &               &   0.66$\pm$4.71 \\
08 &   10.39$\pm$0.04  &               &   8.20$\pm$5.51 \\
09 &    7.30$\pm$0.02  &               &   6.25$\pm$5.66 \\
10 &   18.44$\pm$0.04  &               &   7.09$\pm$4.72 \\
\hline                                 
\end{tabular}
\begin{tablenotes}
\small
\item The RM column shows the RM measurements from this work.
\item The RM$_{\rm NVSS}$ column shows the RM measurements from \citet{taylor2009}.
\item The GRM column shows the GRM measurements from \citet{hutschenreuter2020}.
\end{tablenotes}
\end{table}

\subsection{Morphology and polarized emission}

By inspecting the 6$^{\prime\prime}$ total-intensity images of our detected polarized sources (see Fig.~\ref{fig:cutouts}), we identify three compact sources (sources 01, 03, and 04, classified as blazars in NED), three FR~II radio galaxies (sources 06, 07, and 09), and three FR~I radio galaxies (sources 05, 08, and 10). Source 02 seems to be a double-lobed radio galaxy. The number of FR~I radio galaxies detected in polarization represents a much greater fraction of this type of object than seen in the first data release area of LoTSS (\citealt{vaneck2018} found one out of 92 detected polarized sources), which suggests that more sensitive observations may help with the detection of the polarized emission from extended regions of FR~I radio galaxies \citep{osullivan2018b}. Sources 01, 03, and 04 have been classified as BL Lacertae objects by \citet{healey2008}, \citet{abdo2010}, and \citet{dabrusco2014}, respectively. Future observations of these sources (and in particular of source 02) with the addition of the LOFAR international baselines (which could provide an angular resolution of 0.4$^{\prime\prime}$ at 140~MHz, see \citealt{moldon2015}) would reveal the source morphology in detail.

The FR\,I-type sources have the lowest mean redshift (0.1470), compared to the blazars (0.3733) and the FR\,II sources (0.7062). This is expected and consistent with the general trend in radio power for FR\,I versus FR\,II sources (see e.g., \citealt{mingo2019}). However, the FR\,I-type sources have the largest absolute RMs (16.3~rad\,m$^{-2}$), compared to the blazars (6.7~rad\,m$^{-2}$) and the FR\,II sources (5.0~rad\,m$^{-2}$). This may be reflecting the environment in which the polarized emission originates, for example the inner jet for FR\,I sources versus hotspots of FR\,II.

Some of our detected polarized sources show polarization in their core or inner jets. Since the core of a radio galaxy is not expected to be strongly polarized (e.g., \citealt{saikia1998}), this may represent polarized emission originating from the inner jets in a region of highly ordered magnetic field, or possibly indicate restarted radio galaxy activity (e.g., \citealt{mahatma2019}).

In Fig.~\ref{fig:cutouts}, we can also see that \citet{grant2010} detect one of the lobes of source 07 at 1.4~GHz that is not detected using LOFAR at 146~MHz. This may be an example of the Laing-Garrington effect \citep{laing1988,garrington1988}, which is based on the fact that depolarization with increasing wavelength is usually larger for the weaker flux density lobe (corresponding to the counter-jet farther from us). In principle, large numbers of detections of this effect can probe the properties of the environments of distant radio galaxies.

\section{Summary}
\label{sec:con}

We used LOFAR to observe the ELAIS-N1 field at 146 MHz over several epochs and analyzed the resulting 20$^{\prime\prime}$-resolution polarimetric images. In order to achieve sensitivities not limited by the thermal noise of standard eight-hour LOFAR observations, a stacking technique was developed. The outcomes of this analysis are:

   \begin{enumerate}
      \item We have demonstrated the feasibility of the stacking technique presented here. With this technique, we were able to reach a 1$\sigma_{\rm QU}$ rms noise level of 26~$\mu$Jy~beam$^{-1}$ in the central part of the field. 
      After stacking six epochs, the median noise across the field was reduced by $\sim$\!$\sqrt{6}$ in comparison to the reference epoch.
      \item We detected ten polarized sources in the stacked data, using an 8$\sigma_{\rm QU}$ detection threshold. Seven of these sources were not detected in the reference epoch alone. We have presented radio images and RM measurements of the detected sources.
      \item Since the imaged area is of 16~deg$^2$, the polarized source surface density is one per 1.6~deg$^2$ at the resolution and sensitivity of our observations. This should be considered a lower limit of the true number density.
      \item We have detected a larger fraction of FR~I radio galaxies (three out of ten) than in previous, less sensitive observations (e.g., one out of 92, \citealt{vaneck2018}), which suggests that increasingly sensitive LOFAR observations may help to detect and characterize the magnetic field structure in this type of object at low frequencies.
   \end{enumerate}

This work provides a valuable tool for the study of deep observations in polarization at very low frequencies, and in particular for the initial analysis of the ELAIS-N1 field. The next steps will be to combine all of the available observations of the field (including new observations taken in recent cycles and further allocated observations of the field), to further decrease the detection threshold, to characterize the reliability and completeness of the catalog of detected sources, and to analyze the polarized source counts on the basis of an increased number of detected sources. In addition, the stacking technique presented here can also be applied to the other LoTSS Deep Fields (Lockman Hole and Boötes) in the future, as well as to the GOODS-N (Great Observatories Origins Deep Survey - North) field.

\vspace{0.5cm}

\begin{acknowledgements}
      We thank the anonymous referee for the valuable and constructive comments, which have helped to improve this paper. This paper is based on data obtained with the International LOFAR Telescope (ILT) under project codes LC0\_019, LC2\_024 and LC4\_008. LOFAR \citep{vanhaarlem2013} is the Low Frequency Array designed and constructed by ASTRON. It has observing, data processing, and data storage facilities in several countries, that are owned by various parties (each with their own funding sources), and that are collectively operated by the ILT foundation under a joint scientific policy. The ILT resources have benefitted from the following recent major funding sources: CNRS-INSU, Observatoire de Paris and Universit\'{e} d'Orl\'{e}ans, France; BMBF, MIWF-NRW, MPG, Germany; Science Foundation Ireland (SFI), Department of Business, Enterprise and Innovation (DBEI), Ireland; NWO, The Netherlands; The Science and Technology Facilities Council, UK. 
      NHR acknowledges support from the BMBF through projects D-LOFAR IV (FKZ: 05A17PC1) and MeerKAT (FKZ: 05A17PC2). VJ acknowledges support by the Croatian Science Foundation for the project IP-2018-01-2889 (LowFreqCRO). The authors of the Polish scientific institutions thank the Ministry of Science and Higher Education (MSHE), Poland for granting funds for the Polish contribution to the ILT (MSHE decision no. DIR/WK/2016/2017/05-1)" and for maintenance of the LOFAR PL-610 Borowiec, LOFAR PL-611 Lazy, LOFAR PL-612 Baldy stations. JS is grateful for support from the UK STFC via grant ST/R000972/1. MJH acknowledges support from the UK Science and Technology Facilities Council (ST/R000905/1). RK acknowledges support from the Science and Technology Facilities Council (STFC) through an STFC studentship via grant ST/R504737/1. PNB is grateful for support from the UK STFC via grant ST/R000972/1. IP acknowledges support from INAF under the SKA/CTA PRIN “FORECaST” and the PRIN MAIN STREAM “SAuROS” projects.
      This research made use of Montage. It is funded by the National Science Foundation under Grant Number ACI-1440620, and was previously funded by the National Aeronautics and Space Administration's Earth Science Technology Office, Computation Technologies Project, under Cooperative Agreement Number NCC5-626 between NASA and the California Institute of Technology. This research made use of \texttt{Topcat} \citep{taylor2005}, available at \url{http://www.starlink.ac.uk/topcat/}, \texttt{APLpy}, an open-source plotting package for Python hosted at \url{http://aplpy.github.com}, and  \texttt{Astropy}, a community-developed core Python package for Astronomy \citep{astropy2013}. This research made use of the NASA/IPAC Extragalactic Database (NED), which is operated by the Jet Propulsion Laboratory, California Institute of Technology, under contract with the National Aeronautics and Space Administration.
\end{acknowledgements}

%-------------------------------------------------------------------

\setlength{\bibsep}{0pt plus 0.3ex}
\bibliographystyle{agsm}
\bibliography{references}

@ARTICLE{abdo2010,
       author = {{Abdo}, A.~A. and {Ackermann}, M. and {Ajello}, M. and others},
        title = {},
      journal = {\apj},
         year = "2010",
        month = "May",
       volume = {715},
       number = {1},
        pages = {429-457},
          doi = {10.1088/0004-637X/715/1/429},
archivePrefix = {arXiv},
       eprint = {1002.0150},
 primaryClass = {astro-ph.HE},
       adsurl = {https://ui.adsabs.harvard.edu/abs/2010ApJ...715..429A}
}

@ARTICLE{aihara2018,
       author = {{Aihara}, Hiroaki and {Armstrong}, Robert and {Bickerton}, Steven and others},
        title = {},
      journal = {\pasj},
         year = 2018,
        month = jan,
       volume = {70},
          eid = {S8},
        pages = {S8},
          doi = {10.1093/pasj/psx081},
archivePrefix = {arXiv},
       eprint = {1702.08449},
 primaryClass = {astro-ph.IM},
       adsurl = {https://ui.adsabs.harvard.edu/abs/2018PASJ...70S...8A}
}

@ARTICLE{albareti2017,
       author = {{Albareti}, Franco D. and {Allende Prieto}, Carlos and
         {Almeida}, Andres and others},
        title = {},
      journal = {\apjs},
         year = 2017,
        month = dec,
       volume = {233},
       number = {2},
          eid = {25},
        pages = {25},
          doi = {10.3847/1538-4365/aa8992},
archivePrefix = {arXiv},
       eprint = {1608.02013},
 primaryClass = {astro-ph.GA},
       adsurl = {https://ui.adsabs.harvard.edu/abs/2017ApJS..233...25A}
}

@ARTICLE{anderson2015,
       author = {{Anderson}, C.~S. and {Gaensler}, B.~M. and {Feain}, I.~J. and
         {Franzen}, T.~M.~O.},
        title = {},
      journal = {\apj},
         year = 2015,
        month = dec,
       volume = {815},
       number = {1},
          eid = {49},
        pages = {49},
          doi = {10.1088/0004-637X/815/1/49},
archivePrefix = {arXiv},
       eprint = {1511.04080},
primaryClass = {astro-ph.GA},
       adsurl = {https://ui.adsabs.harvard.edu/abs/2015ApJ...815...49A}
}

@ARTICLE{anderson2019,
       author = {{Anderson}, C.~S. and {O'Sullivan}, S.~P. and {Heald}, G.~H. and
         {Hodgson}, T. and {Pasetto}, A. and {Gaensler}, B.~M.},
        title = {},
      journal = {\mnras},
         year = 2019,
        month = may,
       volume = {485},
       number = {3},
        pages = {3600-3622},
          doi = {10.1093/mnras/stz377},
archivePrefix = {arXiv},
       eprint = {1901.08066},
 primaryClass = {astro-ph.GA},
       adsurl = {https://ui.adsabs.harvard.edu/abs/2019MNRAS.485.3600A}
}

@ARTICLE{astropy2013,
   author = {{Astropy Collaboration} and {Robitaille}, T.~P. and {Tollerud}, E.~J. and 
	{Greenfield}, P. and others},
    title = {},
  journal = {\aap},
archivePrefix = "arXiv",
   eprint = {1307.6212},
 primaryClass = "astro-ph.IM",
     year = 2013,
    month = oct,
   volume = 558,
      eid = {A33},
    pages = {A33},
      doi = {10.1051/0004-6361/201322068},
   adsurl = {http://adsabs.harvard.edu/abs/2013A%26A...558A..33A}
}

@INBOOK{beck2013,
       author = {{Beck}, Rainer and {Wielebinski}, Richard},
        title = {},
    booktitle = {Planets, Stars and Stellar Systems. Volume 5: Galactic Structure and Stellar Populations},
         year = "2013",
       editor = {{Oswalt}, Terry D. and {Gilmore}, Gerard},
       volume = {5},
        pages = {641},
          doi = {10.1007/978-94-007-5612-0_13},
       adsurl = {https://ui.adsabs.harvard.edu/abs/2013pss5.book..641B}
}

@ARTICLE{becker1995,
       author = {{Becker}, Robert H. and {White}, Richard L. and {Helfand}, David J.},
        title = {},
      journal = {\apj},
         year = "1995",
        month = "Sep",
       volume = {450},
        pages = {559},
          doi = {10.1086/176166},
       adsurl = {https://ui.adsabs.harvard.edu/abs/1995ApJ...450..559B}
}

@ARTICLE{bernardi2013,
       author = {{Bernardi}, G. and {Greenhill}, L.~J. and {Mitchell}, D.~A. and others},
        title = {},
      journal = {\apj},
         year = "2013",
        month = "Jul",
       volume = {771},
       number = {2},
          eid = {105},
        pages = {105},
          doi = {10.1088/0004-637X/771/2/105},
archivePrefix = {arXiv},
       eprint = {1305.6047},
 primaryClass = {astro-ph.CO},
       adsurl = {https://ui.adsabs.harvard.edu/abs/2013ApJ...771..105B}
}

@INBOOK{berriman2003,
       author = {{Berriman}, G.~B. and {Good}, J.~C. and {Curkendall}, D.~W. and others},
        title = {},
    booktitle = {Astronomical Data Analysis Software and Systems XII ASP Conference Series, Vol. 295, 2003 H. E. Payne, R. I. Jedrzejewski, and R. N. Hook, eds., p.343},
         year = "2003",
       editor = {{Payne}, H.~E. and {Jedrzejewski}, R.~I. and {Hook}, R.~N.},
       volume = {295},
       series = {Astronomical Society of the Pacific Conference Series},
        pages = {343},
       adsurl = {https://ui.adsabs.harvard.edu/abs/2003ASPC..295..343B}
}

@ARTICLE{bonafede2010,
       author = {{Bonafede}, A. and {Feretti}, L. and {Murgia}, M. and others},
        title = {},
      journal = {\aap},
         year = 2010,
        month = apr,
       volume = {513},
          eid = {A30},
        pages = {A30},
          doi = {10.1051/0004-6361/200913696},
archivePrefix = {arXiv},
       eprint = {1002.0594},
 primaryClass = {astro-ph.CO},
       adsurl = {https://ui.adsabs.harvard.edu/abs/2010A&A...513A..30B}
}

@ARTICLE{brentjens2005,
       author = {{Brentjens}, M.~A. and {de Bruyn}, A.~G.},
        title = {},
      journal = {\aap},
         year = "2005",
        month = "Oct",
       volume = {441},
       number = {3},
        pages = {1217-1228},
          doi = {10.1051/0004-6361:20052990},
archivePrefix = {arXiv},
       eprint = {astro-ph/0507349},
 primaryClass = {astro-ph},
       adsurl = {https://ui.adsabs.harvard.edu/abs/2005A&A...441.1217B}
}

@ARTICLE{burn1966,
       author = {{Burn}, B.~J.},
        title = {},
      journal = {\mnras},
         year = "1966",
        month = "Jan",
       volume = {133},
        pages = {67},
          doi = {10.1093/mnras/133.1.67},
       adsurl = {https://ui.adsabs.harvard.edu/abs/1966MNRAS.133...67B}
}

@ARTICLE{chakraborty2019,
       author = {{Chakraborty}, Arnab and {Roy}, Nirupam and {Datta}, Abhirup and others},
        title = {},
      journal = {\mnras},
         year = 2019,
        month = nov,
       volume = {490},
       number = {1},
        pages = {243-259},
          doi = {10.1093/mnras/stz2533},
archivePrefix = {arXiv},
       eprint = {1908.10380},
 primaryClass = {astro-ph.CO},
       adsurl = {https://ui.adsabs.harvard.edu/abs/2019MNRAS.490..243C}
}

@ARTICLE{condon1998,
       author = {{Condon}, J.~J. and {Cotton}, W.~D. and {Greisen}, E.~W. and others},
        title = {},
      journal = {\aj},
         year = "1998",
        month = "May",
       volume = {115},
       number = {5},
        pages = {1693-1716},
          doi = {10.1086/300337},
       adsurl = {https://ui.adsabs.harvard.edu/abs/1998AJ....115.1693C}
}

@ARTICLE{condon2012,
       author = {{Condon}, J.~J. and {Cotton}, W.~D. and {Fomalont}, E.~B. and others},
        title = {},
      journal = {\apj},
         year = 2012,
        month = oct,
       volume = {758},
       number = {1},
          eid = {23},
        pages = {23},
          doi = {10.1088/0004-637X/758/1/23},
archivePrefix = {arXiv},
       eprint = {1207.2439},
 primaryClass = {astro-ph.CO},
       adsurl = {https://ui.adsabs.harvard.edu/abs/2012ApJ...758...23C}
}

@ARTICLE{dabrusco2014,
       author = {{D'Abrusco}, R. and {Massaro}, F. and {Paggi}, A. and others},
        title = {},
      journal = {\apjs},
         year = "2014",
        month = "Nov",
       volume = {215},
       number = {1},
          eid = {14},
        pages = {14},
          doi = {10.1088/0067-0049/215/1/14},
archivePrefix = {arXiv},
       eprint = {1410.0029},
 primaryClass = {astro-ph.HE},
       adsurl = {https://ui.adsabs.harvard.edu/abs/2014ApJS..215...14D}
}

@ARTICLE{degasperin2019,
       author = {{de Gasperin}, F. and {Dijkema}, T.~J. and {Drabent}, A. and others},
        title = {},
      journal = {\aap},
         year = "2019",
        month = "Feb",
       volume = {622},
          eid = {A5},
        pages = {A5},
          doi = {10.1051/0004-6361/201833867},
archivePrefix = {arXiv},
       eprint = {1811.07954},
 primaryClass = {astro-ph.IM},
       adsurl = {https://ui.adsabs.harvard.edu/abs/2019A&A...622A...5D}
}

@ARTICLE{duncan2020,
       author = {{Duncan}, K. and others},
        title = {},
      journal = {\aap},
         year = "2020"
}

@ARTICLE{garrington1988,
       author = {{Garrington}, S.~T. and {Leahy}, J.~P. and {Conway}, R.~G. and
         {Laing}, R.~A.},
        title = {},
      journal = {\nat},
         year = "1988",
        month = "Jan",
       volume = {331},
       number = {6152},
        pages = {147-149},
          doi = {10.1038/331147a0},
       adsurl = {https://ui.adsabs.harvard.edu/abs/1988Natur.331..147G}
}

@ARTICLE{falco1998,
       author = {{Falco}, E.~E. and {Kochanek}, C.~S. and {Mu{\~n}oz}, J.~A.},
        title = {},
      journal = {\apj},
         year = 1998,
        month = feb,
       volume = {494},
       number = {1},
        pages = {47-59},
          doi = {10.1086/305207},
archivePrefix = {arXiv},
       eprint = {astro-ph/9707032},
 primaryClass = {astro-ph},
       adsurl = {https://ui.adsabs.harvard.edu/abs/1998ApJ...494...47F}
}

@ARTICLE{farnes2014,
       author = {{Farnes}, J.~S. and {Gaensler}, B.~M. and {Carretti}, E.},
        title = {},
      journal = {\apjs},
         year = 2014,
        month = may,
       volume = {212},
       number = {1},
          eid = {15},
        pages = {15},
          doi = {10.1088/0067-0049/212/1/15},
archivePrefix = {arXiv},
       eprint = {1403.2391},
 primaryClass = {astro-ph.GA},
       adsurl = {https://ui.adsabs.harvard.edu/abs/2014ApJS..212...15F}
}

@ARTICLE{george2012,
       author = {{George}, Samuel J. and {Stil}, Jeroen M. and {Keller}, Ben W.},
        title = {},
      journal = {\pasa},
         year = "2012",
        month = "Oct",
       volume = {29},
       number = {3},
        pages = {214-220},
          doi = {10.1071/AS11027},
archivePrefix = {arXiv},
       eprint = {1106.5362},
 primaryClass = {astro-ph.IM},
       adsurl = {https://ui.adsabs.harvard.edu/abs/2012PASA...29..214G}
}

@ARTICLE{grant2010,
       author = {{Grant}, J.~K. and {Taylor}, A.~R. and {Stil}, J.~M. and others},
        title = {},
      journal = {\apj},
         year = "2010",
        month = "May",
       volume = {714},
       number = {2},
        pages = {1689-1701},
          doi = {10.1088/0004-637X/714/2/1689},
archivePrefix = {arXiv},
       eprint = {1003.4460},
 primaryClass = {astro-ph.CO},
       adsurl = {https://ui.adsabs.harvard.edu/abs/2010ApJ...714.1689G}
}

@ARTICLE{hales2014,
       author = {{Hales}, C.~A. and others},
        title = {},
      journal = {\mnras},
         year = 2014,
        month = jul,
       volume = {441},
       number = {3},
        pages = {2555-2592},
          doi = {10.1093/mnras/stu576},
archivePrefix = {arXiv},
       eprint = {1403.5307},
 primaryClass = {astro-ph.GA},
       adsurl = {https://ui.adsabs.harvard.edu/abs/2014MNRAS.441.2555H}
}

@ARTICLE{hales2014b,
       author = {{Hales}, C.~A. and others},
        title = {},
      journal = {\mnras},
         year = 2014,
        month = jun,
       volume = {440},
       number = {4},
        pages = {3113-3139},
          doi = {10.1093/mnras/stu500},
archivePrefix = {arXiv},
       eprint = {1403.5308},
 primaryClass = {astro-ph.GA},
       adsurl = {https://ui.adsabs.harvard.edu/abs/2014MNRAS.440.3113H}
}

@ARTICLE{Han2017,
       author = {{Han}, J.~L.},
        title = {},
      journal = {\araa},
         year = "2017",
        month = "Aug",
       volume = {55},
       number = {1},
        pages = {111-157},
          doi = {10.1146/annurev-astro-091916-055221},
       adsurl = {https://ui.adsabs.harvard.edu/abs/2017ARA&A..55..111H}
}

@ARTICLE{hardcastle1997,
       author = {{Hardcastle}, M.~J. and {Alexander}, P. and {Pooley}, G.~G. and
         {Riley}, J.~M.},
        title = {},
      journal = {\mnras},
         year = 1997,
        month = jul,
       volume = {288},
       number = {4},
        pages = {859-890},
          doi = {10.1093/mnras/288.4.859},
       adsurl = {https://ui.adsabs.harvard.edu/abs/1997MNRAS.288..859H}
}

@ARTICLE{heald2009,
       author = {{Heald}, G. and {Braun}, R. and {Edmonds}, R.},
        title = {},
      journal = {\aap},
         year = 2009,
        month = aug,
       volume = {503},
       number = {2},
        pages = {409-435},
          doi = {10.1051/0004-6361/200912240},
archivePrefix = {arXiv},
       eprint = {0905.3995},
 primaryClass = {astro-ph.GA},
       adsurl = {https://ui.adsabs.harvard.edu/abs/2009A&A...503..409H}
}

@ARTICLE{heald2020,
       author = {{Heald}, George and {Mao}, Sui and {Vacca}, Valentina and others},
        title = {},
      journal = {Galaxies},
         year = 2020,
        month = jul,
       volume = {8},
       number = {3},
        pages = {53},
          doi = {10.3390/galaxies8030053},
archivePrefix = {arXiv},
       eprint = {2006.03172},
 primaryClass = {astro-ph.GA},
       adsurl = {https://ui.adsabs.harvard.edu/abs/2020Galax...8...53H}
}

@ARTICLE{healey2008,
       author = {{Healey}, Stephen E. and {Romani}, Roger W. and {Cotter}, Garret and others},
        title = {},
      journal = {\apjs},
         year = "2008",
        month = "Mar",
       volume = {175},
       number = {1},
        pages = {97-104},
          doi = {10.1086/523302},
archivePrefix = {arXiv},
       eprint = {0709.1735},
 primaryClass = {astro-ph},
       adsurl = {https://ui.adsabs.harvard.edu/abs/2008ApJS..175...97H}
}

@ARTICLE{hopkins2003,
       author = {{Hopkins}, A.~M. and {Afonso}, J. and {Chan}, B. and others},
        title = {},
      journal = {\aj},
         year = "2003",
        month = "Feb",
       volume = {125},
       number = {2},
        pages = {465-477},
          doi = {10.1086/345974},
archivePrefix = {arXiv},
       eprint = {astro-ph/0211068},
 primaryClass = {astro-ph},
       adsurl = {https://ui.adsabs.harvard.edu/abs/2003AJ....125..465H}
}

@ARTICLE{hurley-walker2017,
       author = {{Hurley-Walker}, N. and {Callingham}, J.~R. and {Hancock}, P.~J. and others},
        title = {},
      journal = {\mnras},
         year = "2017",
        month = "Jan",
       volume = {464},
       number = {1},
        pages = {1146-1167},
          doi = {10.1093/mnras/stw2337},
archivePrefix = {arXiv},
       eprint = {1610.08318},
 primaryClass = {astro-ph.GA},
       adsurl = {https://ui.adsabs.harvard.edu/abs/2017MNRAS.464.1146H}
}

@ARTICLE{hutschenreuter2020,
       author = {{Hutschenreuter}, Sebastian and {En{\ss}lin}, Torsten A.},
        title = {},
      journal = {\aap},
         year = "2020",
        month = "Jan",
       volume = {633},
          eid = {A150},
        pages = {A150},
          doi = {10.1051/0004-6361/201935479},
archivePrefix = {arXiv},
       eprint = {1903.06735},
 primaryClass = {astro-ph.GA},
       adsurl = {https://ui.adsabs.harvard.edu/abs/2020A&A...633A.150H}
}

@ARTICLE{intema2009,
       author = {{Intema}, H.~T. and {van der Tol}, S. and {Cotton}, W.~D. and others},
        title = {},
      journal = {\aap},
         year = "2009",
        month = "Jul",
       volume = {501},
       number = {3},
        pages = {1185-1205},
          doi = {10.1051/0004-6361/200811094},
archivePrefix = {arXiv},
       eprint = {0904.3975},
 primaryClass = {astro-ph.IM},
       adsurl = {https://ui.adsabs.harvard.edu/abs/2009A&A...501.1185I}
}

@ARTICLE{jelic2014,
       author = {{Jeli{\'c}}, V. and {de Bruyn}, A.~G. and {Mevius}, M. and others},
        title = {},
      journal = {\aap},
         year = "2014",
        month = "Aug",
       volume = {568},
          eid = {A101},
        pages = {A101},
          doi = {10.1051/0004-6361/201423998},
archivePrefix = {arXiv},
       eprint = {1407.2093},
 primaryClass = {astro-ph.GA},
       adsurl = {https://ui.adsabs.harvard.edu/abs/2014A&A...568A.101J}
}

@ARTICLE{kessler1996,
       author = {{Kessler}, M.~F. and {Steinz}, J.~A. and {Anderegg}, M.~E. and others},
        title = {},
      journal = {\aap},
         year = "1996",
        month = "Nov",
       volume = {500},
        pages = {493-497},
       adsurl = {https://ui.adsabs.harvard.edu/abs/1996A&A...315L..27K}
}

@ARTICLE{kondapally2020,
       author = {{Kondapally}, R. and others},
        title = {},
      journal = {\aap},
         year = "2020"
}

@ARTICLE{laing1988,
       author = {{Laing}, R.~A.},
        title = {},
      journal = {\nat},
         year = "1988",
        month = "Jan",
       volume = {331},
       number = {6152},
        pages = {149-151},
          doi = {10.1038/331149a0},
       adsurl = {https://ui.adsabs.harvard.edu/abs/1988Natur.331..149L}
}

@ARTICLE{laing2014,
       author = {{Laing}, R.~A. and {Bridle}, A.~H.},
        title = {},
      journal = {\mnras},
         year = 2014,
        month = feb,
       volume = {437},
       number = {4},
        pages = {3405-3441},
          doi = {10.1093/mnras/stt2138},
archivePrefix = {arXiv},
       eprint = {1311.1015},
 primaryClass = {astro-ph.CO},
       adsurl = {https://ui.adsabs.harvard.edu/abs/2014MNRAS.437.3405L}
}

@ARTICLE{lawrence2007,
       author = {{Lawrence}, A. and {Warren}, S.~J. and {Almaini}, O. and others},
        title = {},
      journal = {\mnras},
         year = 2007,
        month = aug,
       volume = {379},
       number = {4},
        pages = {1599-1617},
          doi = {10.1111/j.1365-2966.2007.12040.x},
archivePrefix = {arXiv},
       eprint = {astro-ph/0604426},
 primaryClass = {astro-ph},
       adsurl = {https://ui.adsabs.harvard.edu/abs/2007MNRAS.379.1599L}
}

@ARTICLE{lenc2016,
       author = {{Lenc}, E. and {Gaensler}, B.~M. and {Sun}, X.~H. and others},
        title = {},
      journal = {\apj},
         year = "2016",
        month = "Oct",
       volume = {830},
       number = {1},
          eid = {38},
        pages = {38},
          doi = {10.3847/0004-637X/830/1/38},
archivePrefix = {arXiv},
       eprint = {1607.05779},
 primaryClass = {astro-ph.GA},
       adsurl = {https://ui.adsabs.harvard.edu/abs/2016ApJ...830...38L}
}

@ARTICLE{lonsdale2003,
       author = {{Lonsdale}, Carol J. and {Smith}, Harding E. and
         {Rowan-Robinson}, Michael and others},
        title = {},
      journal = {\pasp},
         year = 2003,
        month = aug,
       volume = {115},
       number = {810},
        pages = {897-927},
          doi = {10.1086/376850},
archivePrefix = {arXiv},
       eprint = {astro-ph/0305375},
 primaryClass = {astro-ph},
       adsurl = {https://ui.adsabs.harvard.edu/abs/2003PASP..115..897L}
}

@ARTICLE{mahatma2019,
       author = {{Mahatma}, V.~H. and {Hardcastle}, M.~J. and {Williams}, W.~L. and others},
        title = {},
      journal = {\aap},
         year = "2019",
        month = "Feb",
       volume = {622},
          eid = {A13},
        pages = {A13},
          doi = {10.1051/0004-6361/201833973},
archivePrefix = {arXiv},
       eprint = {1811.08194},
 primaryClass = {astro-ph.GA},
       adsurl = {https://ui.adsabs.harvard.edu/abs/2019A&A...622A..13M}
}

@ARTICLE{manners2003,
       author = {{Manners}, J.~C. and {Johnson}, O. and {Almaini}, O. and others},
        title = {},
      journal = {\mnras},
         year = 2003,
        month = jul,
       volume = {343},
       number = {1},
        pages = {293-305},
          doi = {10.1046/j.1365-8711.2003.06672.x},
archivePrefix = {arXiv},
       eprint = {astro-ph/0207622},
 primaryClass = {astro-ph},
       adsurl = {https://ui.adsabs.harvard.edu/abs/2003MNRAS.343..293M}
}

@ARTICLE{mao2014,
       author = {{Mao}, Sui Ann and {Banfield}, Julie and {Gaensler}, Bryan and others},
        title = {},
      journal = {arXiv e-prints},
         year = 2014,
        month = jan,
          eid = {arXiv:1401.1875},
        pages = {arXiv:1401.1875},
archivePrefix = {arXiv},
       eprint = {1401.1875},
 primaryClass = {astro-ph.GA},
       adsurl = {https://ui.adsabs.harvard.edu/abs/2014arXiv1401.1875M}
}

@ARTICLE{martin2005,
       author = {{Martin}, D. Christopher and {Fanson}, James and {Schiminovich}, David and others},
        title = {},
      journal = {\apjl},
         year = 2005,
        month = jan,
       volume = {619},
       number = {1},
        pages = {L1-L6},
          doi = {10.1086/426387},
archivePrefix = {arXiv},
       eprint = {astro-ph/0411302},
 primaryClass = {astro-ph},
       adsurl = {https://ui.adsabs.harvard.edu/abs/2005ApJ...619L...1M}
}

@ARTICLE{mauduit2012,
       author = {{Mauduit}, J. -C. and {Lacy}, M. and {Farrah}, D. and others},
        title = {},
      journal = {\pasp},
         year = 2012,
        month = jul,
       volume = {124},
       number = {917},
        pages = {714},
          doi = {10.1086/666945},
archivePrefix = {arXiv},
       eprint = {1206.4060},
 primaryClass = {astro-ph.CO},
       adsurl = {https://ui.adsabs.harvard.edu/abs/2012PASP..124..714M}
}

@ARTICLE{mcmahon2001,
       author = {{McMahon}, R.~G. and {Walton}, N.~A. and {Irwin}, M.~J. and others},
        title = {},
      journal = {\nar},
         year = "2001",
        month = "Jan",
       volume = {45},
       number = {1-2},
        pages = {97-104},
          doi = {10.1016/S1387-6473(00)00137-8},
archivePrefix = {arXiv},
       eprint = {astro-ph/0001285},
 primaryClass = {astro-ph},
       adsurl = {https://ui.adsabs.harvard.edu/abs/2001NewAR..45...97M}
}

@MISC{mevius2018,
       author = {{Mevius}, Maaijke},
        title = "{RMextract: Ionospheric Faraday Rotation calculator}",
         year = "2018",
        month = "Jun",
          eid = {ascl:1806.024},
        pages = {ascl:1806.024},
archivePrefix = {ascl},
       eprint = {1806.024},
       adsurl = {https://ui.adsabs.harvard.edu/abs/2018ascl.soft06024M}
}

@ARTICLE{michilli2018,
       author = {{Michilli}, D. and {Seymour}, A. and {Hessels}, J.~W.~T. and others},
        title = {},
      journal = {\nat},
         year = 2018,
        month = jan,
       volume = {553},
       number = {7687},
        pages = {182-185},
          doi = {10.1038/nature25149},
archivePrefix = {arXiv},
       eprint = {1801.03965},
 primaryClass = {astro-ph.HE},
       adsurl = {https://ui.adsabs.harvard.edu/abs/2018Natur.553..182M}
}

@ARTICLE{mingo2019,
       author = {{Mingo}, B. and {Croston}, J.~H. and {Hardcastle}, M.~J. and others},
        title = {},
      journal = {\mnras},
         year = 2019,
        month = sep,
       volume = {488},
       number = {2},
        pages = {2701-2721},
          doi = {10.1093/mnras/stz1901},
archivePrefix = {arXiv},
       eprint = {1907.03726},
 primaryClass = {astro-ph.GA},
       adsurl = {https://ui.adsabs.harvard.edu/abs/2019MNRAS.488.2701M}
}

@MISC{mohan2015,
       author = {{Mohan}, Niruj and {Rafferty}, David},
        title = {},
         year = 2015,
        month = feb,
          eid = {ascl:1502.007},
        pages = {ascl:1502.007},
archivePrefix = {ascl},
       eprint = {1502.007},
       adsurl = {https://ui.adsabs.harvard.edu/abs/2015ascl.soft02007M}
}

@ARTICLE{moldon2015,
       author = {{Mold{\'o}n}, J. and {Deller}, A.~T. and {Wucknitz}, O. and others},
        title = {},
      journal = {\aap},
         year = 2015,
        month = feb,
       volume = {574},
          eid = {A73},
        pages = {A73},
          doi = {10.1051/0004-6361/201425042},
archivePrefix = {arXiv},
       eprint = {1411.2743},
 primaryClass = {astro-ph.IM},
       adsurl = {https://ui.adsabs.harvard.edu/abs/2015A&A...574A..73M}
}

@ARTICLE{morrissey2007,
       author = {{Morrissey}, Patrick and {Conrow}, Tim and {Barlow}, Tom A. and others},
        title = {},
      journal = {\apjs},
         year = 2007,
        month = dec,
       volume = {173},
       number = {2},
        pages = {682-697},
          doi = {10.1086/520512},
       adsurl = {https://ui.adsabs.harvard.edu/abs/2007ApJS..173..682M}
}

@ARTICLE{mulcahy2014,
       author = {{Mulcahy}, D.~D. and {Horneffer}, A. and {Beck}, R. and others},
        title = {},
      journal = {\aap},
         year = "2014",
        month = "Aug",
       volume = {568},
          eid = {A74},
        pages = {A74},
          doi = {10.1051/0004-6361/201424187},
archivePrefix = {arXiv},
       eprint = {1407.1312},
 primaryClass = {astro-ph.GA},
       adsurl = {https://ui.adsabs.harvard.edu/abs/2014A&A...568A..74M}
}

@ARTICLE{neld2018,
       author = {{Neld}, A. and {Horellou}, C. and {Mulcahy}, D.~D. and others},
        title = {},
      journal = {\aap},
         year = "2018",
        month = "Oct",
       volume = {617},
          eid = {A136},
        pages = {A136},
          doi = {10.1051/0004-6361/201732157},
archivePrefix = {arXiv},
       eprint = {1807.09240},
 primaryClass = {astro-ph.GA},
       adsurl = {https://ui.adsabs.harvard.edu/abs/2018A&A...617A.136N}
}

@ARTICLE{ocran2020,
       author = {{Ocran}, E.~F. and {Taylor}, A.~R. and {Vaccari}, M. and {Ishwara-Chand
        ra}, C.~H. and {Prandoni}, I.},
        title = {},
      journal = {\mnras},
         year = 2020,
        month = jan,
       volume = {491},
       number = {1},
        pages = {1127-1145},
          doi = {10.1093/mnras/stz2954},
archivePrefix = {arXiv},
       eprint = {1910.08355},
 primaryClass = {astro-ph.GA},
       adsurl = {https://ui.adsabs.harvard.edu/abs/2020MNRAS.491.1127O}
}

@ARTICLE{offringa2012,
       author = {{Offringa}, A.~R. and {van de Gronde}, J.~J. and {Roerdink}, J.~B.~T.~M.},
        title = {},
      journal = {\aap},
         year = "2012",
        month = "Mar",
       volume = {539},
          eid = {A95},
        pages = {A95},
          doi = {10.1051/0004-6361/201118497},
archivePrefix = {arXiv},
       eprint = {1201.3364},
 primaryClass = {astro-ph.IM},
       adsurl = {https://ui.adsabs.harvard.edu/abs/2012A&A...539A..95O}
}

@ARTICLE{oliver2000,
       author = {{Oliver}, Seb and {Rowan-Robinson}, Michael and {Alexander}, D.~M. and others},
        title = {},
      journal = {\mnras},
         year = "2000",
        month = "Aug",
       volume = {316},
       number = {4},
        pages = {749-767},
          doi = {10.1046/j.1365-8711.2000.03550.x},
archivePrefix = {arXiv},
       eprint = {astro-ph/0003263},
 primaryClass = {astro-ph},
       adsurl = {https://ui.adsabs.harvard.edu/abs/2000MNRAS.316..749O}
}

@ARTICLE{oliver2012,
       author = {{Oliver}, S.~J. and {Bock}, J. and {Altieri}, B. and others},
        title = {},
      journal = {\mnras},
         year = 2012,
        month = aug,
       volume = {424},
       number = {3},
        pages = {1614-1635},
          doi = {10.1111/j.1365-2966.2012.20912.x},
archivePrefix = {arXiv},
       eprint = {1203.2562},
 primaryClass = {astro-ph.CO},
       adsurl = {https://ui.adsabs.harvard.edu/abs/2012MNRAS.424.1614O}
}

@ARTICLE{oppermann2015,
       author = {{Oppermann}, N. and {Junklewitz}, H. and {Greiner}, M. and others},
        title = {},
      journal = {\aap},
         year = 2015,
        month = mar,
       volume = {575},
          eid = {A118},
        pages = {A118},
          doi = {10.1051/0004-6361/201423995},
archivePrefix = {arXiv},
       eprint = {1404.3701},
 primaryClass = {astro-ph.IM},
       adsurl = {https://ui.adsabs.harvard.edu/abs/2015A&A...575A.118O}
}

@ARTICLE{osullivan2018b,
       author = {{O'Sullivan}, Shane and others},
        title = {},
      journal = {Galaxies},
         year = "2018",
        month = "Nov",
       volume = {6},
       number = {4},
        pages = {126},
          doi = {10.3390/galaxies6040126},
archivePrefix = {arXiv},
       eprint = {1811.12732},
 primaryClass = {astro-ph.HE},
       adsurl = {https://ui.adsabs.harvard.edu/abs/2018Galax...6..126O}
}

@ARTICLE{osullivan2019,
       author = {{O'Sullivan}, S.~P. and {Machalski}, J. and {Van Eck}, C.~L. and others},
        title = {},
      journal = {\aap},
         year = "2019",
        month = "Feb",
       volume = {622},
          eid = {A16},
        pages = {A16},
          doi = {10.1051/0004-6361/201833832},
archivePrefix = {arXiv},
       eprint = {1811.07934},
 primaryClass = {astro-ph.HE},
       adsurl = {https://ui.adsabs.harvard.edu/abs/2019A&A...622A..16O}
}

@ARTICLE{osullivan2020,
       author = {{O'Sullivan}, S.~P. and {Br{\"u}ggen}, M. and {Vazza}, F. and others},
        title = {},
      journal = {\mnras},
         year = 2020,
        month = may,
       volume = {495},
       number = {3},
        pages = {2607-2619},
          doi = {10.1093/mnras/staa1395},
archivePrefix = {arXiv},
       eprint = {2002.06924},
 primaryClass = {astro-ph.CO},
       adsurl = {https://ui.adsabs.harvard.edu/abs/2020MNRAS.495.2607O}
}

@ARTICLE{planck2018,
       author = {{Planck Collaboration} and {Aghanim}, N. and {Akrami}, Y. and others},
        title = {},
      journal = {ArXiv e-prints},
         year = 2018,
        month = jul,
          eid = {arXiv:1807.06209},
        pages = {arXiv:1807.06209},
archivePrefix = {arXiv},
       eprint = {1807.06209},
 primaryClass = {astro-ph.CO},
       adsurl = {https://ui.adsabs.harvard.edu/abs/2018arXiv180706209P}

}

@ARTICLE{rengelink1997,
       author = {{Rengelink}, R.~B. and {Tang}, Y. and {de Bruyn}, A.~G. and others},
        title = {},
      journal = {\aaps},
         year = "1997",
        month = "Aug",
       volume = {124},
        pages = {259-280},
          doi = {10.1051/aas:1997358},
       adsurl = {https://ui.adsabs.harvard.edu/abs/1997A&AS..124..259R}
}

@ARTICLE{richards2009,
       author = {{Richards}, Gordon T. and {Myers}, Adam D. and {Gray}, Alexander G. and others},
        title = {},
      journal = {\apjs},
         year = 2009,
        month = jan,
       volume = {180},
       number = {1},
        pages = {67-83},
          doi = {10.1088/0067-0049/180/1/67},
archivePrefix = {arXiv},
       eprint = {0809.3952},
 primaryClass = {astro-ph},
       adsurl = {https://ui.adsabs.harvard.edu/abs/2009ApJS..180...67R}
}

@ARTICLE{riseley2018,
       author = {{Riseley}, C.~J. and {Lenc}, E. and {Van Eck}, C.~L. and others},
        title = {},
      journal = {\pasa},
         year = "2018",
        month = "Dec",
       volume = {35},
        pages = {43},
          doi = {10.1017/pasa.2018.39},
archivePrefix = {arXiv},
       eprint = {1809.09327},
 primaryClass = {astro-ph.GA},
       adsurl = {https://ui.adsabs.harvard.edu/abs/2018PASA...35...43R}
}

@ARTICLE{riseley2020,
       author = {{Riseley}, C.~J. and {Galvin}, T.~J. and {Sobey}, C. and others},
        title = {},
      journal = {\pasa},
         year = 2020,
        month = jan,
       volume = {37},
          eid = {e029},
        pages = {e029},
          doi = {10.1017/pasa.2020.20},
archivePrefix = {arXiv},
       eprint = {2005.09266},
 primaryClass = {astro-ph.GA},
       adsurl = {https://ui.adsabs.harvard.edu/abs/2020PASA...37...29R}
}

@ARTICLE{sabater2020,
       author = {{Sabater}, P. and others},
        title = {},
      journal = {\aap},
         year = "2020"
}

@ARTICLE{saikia1998,
       author = {{Saikia}, D.~J. and {Kulkarni}, A.~R.},
        title = {},
      journal = {\mnras},
         year = "1998",
        month = "Aug",
       volume = {298},
       number = {3},
        pages = {L45-L48},
          doi = {10.1046/j.1365-8711.1998.01893.x},
       adsurl = {https://ui.adsabs.harvard.edu/abs/1998MNRAS.298L..45S}
}

@ARTICLE{shimwell2017,
       author = {{Shimwell}, T.~W. and {R{\"o}ttgering}, H.~J.~A. and {Best}, P.~N. and others},
        title = {},
      journal = {\aap},
         year = "2017",
        month = "Feb",
       volume = {598},
          eid = {A104},
        pages = {A104},
          doi = {10.1051/0004-6361/201629313},
archivePrefix = {arXiv},
       eprint = {1611.02700},
 primaryClass = {astro-ph.IM},
       adsurl = {https://ui.adsabs.harvard.edu/abs/2017A&A...598A.104S}
}

@ARTICLE{shimwell2019,
       author = {{Shimwell}, T.~W. and {Tasse}, C. and {Hardcastle}, M.~J. and others},
        title = {},
      journal = {\aap},
         year = "2019",
        month = "Feb",
       volume = {622},
          eid = {A1},
        pages = {A1},
          doi = {10.1051/0004-6361/201833559},
archivePrefix = {arXiv},
       eprint = {1811.07926},
 primaryClass = {astro-ph.GA},
       adsurl = {https://ui.adsabs.harvard.edu/abs/2019A&A...622A...1S}
}

@INBOOK{sirothia2009,
       author = {{Sirothia}, S.~K. and {Dennefeld}, M. and {Saikia}, D.~J. and others},
        title = {},
    booktitle = {The Low-Frequency Radio Universe ASP Conference Series, Vol. 407, proceedings of the conference held 8-12 December 2008, at National Centre for Radio Astrophysics (NCRA), TIFR, Pune, India. Edited by D. J. Saikia, D. A. Green, Y. Gupta, and T. Venturi. San Francisco: Astronomical Society of the Pacific, p.27},
         year = "2009",
       editor = {{Saikia}, D.~J. and {Green}, D.~A. and {Gupta}, Y. and {Venturi}, T.},
       volume = {407},
       series = {Astronomical Society of the Pacific Conference Series},
        pages = {27},
       adsurl = {https://ui.adsabs.harvard.edu/abs/2009ASPC..407...27S}
}

@ARTICLE{smirnov2015,
       author = {{Smirnov}, O.~M. and {Tasse}, C.},
        title = {},
      journal = {\mnras},
         year = "2015",
        month = "May",
       volume = {449},
       number = {3},
        pages = {2668-2684},
          doi = {10.1093/mnras/stv418},
archivePrefix = {arXiv},
       eprint = {1502.06974},
 primaryClass = {astro-ph.IM},
       adsurl = {https://ui.adsabs.harvard.edu/abs/2015MNRAS.449.2668S}
}

@ARTICLE{smolcic2017,
       author = {{Smol{\v{c}}i{\'c}}, V. and {Delvecchio}, I. and {Zamorani}, G. and others},
        title = {},
      journal = {\aap},
         year = 2017,
        month = jun,
       volume = {602},
          eid = {A2},
        pages = {A2},
          doi = {10.1051/0004-6361/201630223},
archivePrefix = {arXiv},
       eprint = {1703.09719},
 primaryClass = {astro-ph.GA},
       adsurl = {https://ui.adsabs.harvard.edu/abs/2017A&A...602A...2S}
}

@ARTICLE{sobey2019,
       author = {{Sobey}, C. and {Bilous}, A.~V. and {Grie{\ss}meier}, J. -M. and others},
        title = {},
      journal = {\mnras},
         year = 2019,
        month = apr,
       volume = {484},
       number = {3},
        pages = {3646-3664},
          doi = {10.1093/mnras/stz214},
archivePrefix = {arXiv},
       eprint = {1901.07738},
 primaryClass = {astro-ph.GA},
       adsurl = {https://ui.adsabs.harvard.edu/abs/2019MNRAS.484.3646S}
}

@ARTICLE{sotomayor2013,
       author = {{Sotomayor-Beltran}, C. and {Sobey}, C. and {Hessels}, J.~W.~T. and others},
        title = {},
      journal = {\aap},
         year = "2013",
        month = "Apr",
       volume = {552},
          eid = {A58},
        pages = {A58},
          doi = {10.1051/0004-6361/201220728},
archivePrefix = {arXiv},
       eprint = {1303.6230},
 primaryClass = {astro-ph.IM},
       adsurl = {https://ui.adsabs.harvard.edu/abs/2013A&A...552A..58S}
}

@ARTICLE{sokoloff1998,
       author = {{Sokoloff}, D.~D. and {Bykov}, A.~A. and {Shukurov}, A. and others},
        title = {},
      journal = {\mnras},
         year = 1998,
        month = aug,
       volume = {299},
       number = {1},
        pages = {189-206},
          doi = {10.1046/j.1365-8711.1998.01782.x},
       adsurl = {https://ui.adsabs.harvard.edu/abs/1998MNRAS.299..189S}
}

@ARTICLE{stil2009,
       author = {{Stil}, J.~M. and {Krause}, M. and {Beck}, R. and {Taylor}, A.~R.},
        title = {},
      journal = {\apj},
         year = 2009,
        month = mar,
       volume = {693},
       number = {2},
        pages = {1392-1403},
          doi = {10.1088/0004-637X/693/2/1392},
archivePrefix = {arXiv},
       eprint = {0810.2303},
 primaryClass = {astro-ph},
       adsurl = {https://ui.adsabs.harvard.edu/abs/2009ApJ...693.1392S}
}

@ARTICLE{tasse2014a,
       author = {{Tasse}, Cyril},
        title = {},
      journal = {arXiv e-prints},
         year = "2014",
        month = "Oct",
          eid = {arXiv:1410.8706},
        pages = {arXiv:1410.8706},
archivePrefix = {arXiv},
       eprint = {1410.8706},
 primaryClass = {astro-ph.IM},
       adsurl = {https://ui.adsabs.harvard.edu/abs/2014arXiv1410.8706T}
}

@ARTICLE{tasse2014b,
       author = {{Tasse}, C.},
        title = {},
      journal = {\aap},
         year = "2014",
        month = "Jun",
       volume = {566},
          eid = {A127},
        pages = {A127},
          doi = {10.1051/0004-6361/201423503},
archivePrefix = {arXiv},
       eprint = {1403.6308},
 primaryClass = {astro-ph.IM},
       adsurl = {https://ui.adsabs.harvard.edu/abs/2014A&A...566A.127T}
}

@ARTICLE{tasse2018,
       author = {{Tasse}, C. and {Hugo}, B. and {Mirmont}, M. and others},
        title = {},
      journal = {\aap},
         year = "2018",
        month = "Apr",
       volume = {611},
          eid = {A87},
        pages = {A87},
          doi = {10.1051/0004-6361/201731474},
archivePrefix = {arXiv},
       eprint = {1712.02078},
 primaryClass = {astro-ph.IM},
       adsurl = {https://ui.adsabs.harvard.edu/abs/2018A&A...611A..87T}
}

@ARTICLE{tasse2020,
       author = {{Tasse}, C. and others},
        title = {},
      journal = {\aap},
         year = "2020"
}

@INPROCEEDINGS{taylor2005,
   author = {{Taylor}, M.~B.},
    title = {},
booktitle = {Astronomical Data Analysis Software and Systems XIV},
     year = 2005,
   series = {Astronomical Society of the Pacific Conference Series},
   volume = 347,
   editor = {{Shopbell}, P. and {Britton}, M. and {Ebert}, R.},
    month = dec,
    pages = {29},
   adsurl = {http://adsabs.harvard.edu/abs/2005ASPC..347...29T}
}

@ARTICLE{taylor2007,
       author = {{Taylor}, A.~R. and {Stil}, J.~M. and {Grant}, J.~K. and others},
        title = {},
      journal = {\apj},
         year = "2007",
        month = "Sep",
       volume = {666},
       number = {1},
        pages = {201-211},
          doi = {10.1086/519786},
archivePrefix = {arXiv},
       eprint = {0705.2736},
 primaryClass = {astro-ph},
       adsurl = {https://ui.adsabs.harvard.edu/abs/2007ApJ...666..201T}
}

@ARTICLE{taylor2009,
       author = {{Taylor}, A.~R. and {Stil}, J.~M. and {Sunstrum}, C.},
        title = {},
      journal = {\apj},
         year = 2009,
        month = sep,
       volume = {702},
       number = {2},
        pages = {1230-1236},
          doi = {10.1088/0004-637X/702/2/1230},
       adsurl = {https://ui.adsabs.harvard.edu/abs/2009ApJ...702.1230T}
}

@ARTICLE{tingay2013,
       author = {{Tingay}, S.~J. and {Goeke}, R. and {Bowman}, J.~D. and others},
        title = {},
      journal = {\pasa},
         year = "2013",
        month = "Jan",
       volume = {30},
          eid = {e007},
        pages = {e007},
          doi = {10.1017/pasa.2012.007},
archivePrefix = {arXiv},
       eprint = {1206.6945},
 primaryClass = {astro-ph.IM},
       adsurl = {https://ui.adsabs.harvard.edu/abs/2013PASA...30....7T}
}

@ARTICLE{vacca2016,
       author = {{Vacca}, V. and {Oppermann}, N. and {En{\ss}lin}, T. and others},
        title = {},
      journal = {\aap},
         year = "2016",
        month = "Jun",
       volume = {591},
          eid = {A13},
        pages = {A13},
          doi = {10.1051/0004-6361/201527291},
archivePrefix = {arXiv},
       eprint = {1509.00747},
 primaryClass = {astro-ph.CO},
       adsurl = {https://ui.adsabs.harvard.edu/abs/2016A&A...591A..13V}
}

@ARTICLE{vaneck2011,
       author = {{Van Eck}, C.~L. and {Brown}, J.~C. and {Stil}, J.~M. and others},
        title = {},
      journal = {\apj},
         year = "2011",
        month = "Feb",
       volume = {728},
       number = {2},
          eid = {97},
        pages = {97},
          doi = {10.1088/0004-637X/728/2/97},
archivePrefix = {arXiv},
       eprint = {1012.2938},
 primaryClass = {astro-ph.GA},
       adsurl = {https://ui.adsabs.harvard.edu/abs/2011ApJ...728...97V}
}

@ARTICLE{vaneck2018,
       author = {{Van Eck}, C.~L. and {Haverkorn}, M. and {Alves}, M.~I.~R. and others},
        title = {},
      journal = {\aap},
         year = "2018",
        month = "Jun",
       volume = {613},
          eid = {A58},
        pages = {A58},
          doi = {10.1051/0004-6361/201732228},
archivePrefix = {arXiv},
       eprint = {1801.04467},
 primaryClass = {astro-ph.GA},
       adsurl = {https://ui.adsabs.harvard.edu/abs/2018A&A...613A..58V}
}

@ARTICLE{vanhaarlem2013,
       author = {{van Haarlem}, M.~P. and {Wise}, M.~W. and {Gunst}, A.~W. and others},
        title = {},
      journal = {\aap},
         year = "2013",
        month = "Aug",
       volume = {556},
          eid = {A2},
        pages = {A2},
          doi = {10.1051/0004-6361/201220873},
archivePrefix = {arXiv},
       eprint = {1305.3550},
 primaryClass = {astro-ph.IM},
       adsurl = {https://ui.adsabs.harvard.edu/abs/2013A&A...556A...2V}
}

@ARTICLE{vanweeren2016,
       author = {{van Weeren}, R.~J. and {Williams}, W.~L. and {Hardcastle}, M.~J. and others},
        title = {},
      journal = {\apjs},
         year = "2016",
        month = "Mar",
       volume = {223},
       number = {1},
          eid = {2},
        pages = {2},
          doi = {10.3847/0067-0049/223/1/2},
archivePrefix = {arXiv},
       eprint = {1601.05422},
 primaryClass = {astro-ph.IM},
       adsurl = {https://ui.adsabs.harvard.edu/abs/2016ApJS..223....2V}
}

@ARTICLE{white1997,
       author = {{White}, Richard L. and {Becker}, Robert H. and {Helfand}, David J. and
         {Gregg}, Michael D.},
        title = {},
      journal = {\apj},
         year = "1997",
        month = "Feb",
       volume = {475},
       number = {2},
        pages = {479-493},
          doi = {10.1086/303564},
       adsurl = {https://ui.adsabs.harvard.edu/abs/1997ApJ...475..479W}
}

@ARTICLE{williams2016,
       author = {{Williams}, W.~L. and {van Weeren}, R.~J. and
         {R{\"o}ttgering}, H.~J.~A. and others},
        title = {},
      journal = {\mnras},
         year = "2016",
        month = "Aug",
       volume = {460},
       number = {3},
        pages = {2385-2412},
          doi = {10.1093/mnras/stw1056},
archivePrefix = {arXiv},
       eprint = {1605.01531},
 primaryClass = {astro-ph.CO},
       adsurl = {https://ui.adsabs.harvard.edu/abs/2016MNRAS.460.2385W}
}

\end{document}